\documentclass[final,english]{elsarticle}
\usepackage[T1]{fontenc}
\usepackage[latin9]{inputenc}
\usepackage{varioref}
\usepackage{units}
\usepackage{amsmath}
\usepackage{amssymb}
\usepackage{graphicx}
\usepackage{esint}
\usepackage{caption}
\usepackage{subcaption}
\usepackage{color}

\pdfpageheight\paperheight
\pdfpagewidth\paperwidth

\journal{Elsevier}
\usepackage{upgreek}
\usepackage[bookmarks,bookmarksopen,bookmarksdepth=6]{hyperref}
\usepackage{cases}
\usepackage{url}
\usepackage{lineno}

\usepackage{a4wide}

\usepackage{newtxtext,newtxmath,amsmath}
\makeatother

\usepackage{babel}
\begin{document}

\title{Simulation of the response of SiPMs
       Part~II: with saturation effects }

\author[]{R.~Klanner \corref{cor1}}

\cortext[cor1]{Corresponding author. Email address: Robert.Klanner@desy.de,
 Tel.: +49 40 8998 2558}
\address{ Institute for Experimental Physics, University of Hamburg,
 \\Luruper Chaussee 149, 22761 Hamburg, Germany.}



 \begin{abstract}

 A Monte Carlo program is presented which simulates the response of SiPMs in the nonlinear regime, where the number of Geiger discharges from photons and/or from dark counts in the time interval given by the pulse shape of a single Geiger discharge, approaches or exceeds the number of SiPM pixels.
 The model includes the effects of after-pulses, of prompt and delayed cross-talk, and of the voltage drop over the load resistance of the readout electronics.
 The results of the simulation program are compared to published results from SiPMs with different number of pixels for different intensities and time distributions of photons, dark-count rates, SiPM pulse shapes, and probabilities of cross-talk and after-pulsing.

 \end{abstract}


 \maketitle
 \pagenumbering{arabic}

\section{Introduction}
 \label{sect:Introduction}

 SiPMs (Silicon Photo-Multipliers), matrices of photo-diodes operated above the breakdown voltage, are the photo-detectors of choice for many applications.
 They are robust, have single photon resolution, high photon-detection efficiency, operate at voltages below 100~V and are not affected by magnetic fields.
 However, their detection area is limited, their response is non-linear for high light intensities and high dark-count rates, and their performance is affected by prompt and delayed cross-talk and after-pulses.

 Two operating conditions for SiPMs can be distinguished:
 \begin{enumerate}
   \item \emph{Low occupancy}, if the number of Geiger discharges in a time given by the pulse shape of a single Geiger discharge is small compared to the number of pixels, $N_\mathit{pix}$, and
   \item \emph{High occupancy}, if the number of Geiger discharges approaches or exceeds $N_\mathit{pix}$.
 \end{enumerate}
 In the first case, the response of the SiPM to light pulses is linear.
 The individual Geiger discharges from light and from dark counts, including the effects of cross-talk and after-pulses, can be treated separately and on an event to event basis the current transient and the charge integrated in a gate is calculated.
 This situation is treated in Part~I of the two papers on the simulation of the response of SiPMs~\cite{Garutti:2020}.

 The second case, where the high occupancy of the individual SiPM pixels can be caused by a high number of photons or a high dark-count rate due to background light or radiation damage, is significantly more complicated.
 A simulation program addressing this situation is the topic of this paper.

 Saturation effects for SiPMs are extensively discussed in the literature, and only a short overview will be given.
 Phenomenological parameterisations are discussed in \cite{Grodzicka:2019, Klanner:2019} and references given there.
 Although a large number of different terms and quantities are used in the literature, basically the ratio of the measured mean charge $Q_\mathit{meas}$ to the mean charge of a single Geiger discharge at nominal over-voltage, $Q_1$, as a function of $N_\mathit{seed}$, the mean number of Geiger discharges ignoring effects of non-linearity, cross-talk and after-pulses is parameterised and compared to the experimental data.
 The relation between $N_\mathit{seed}$ and $N_\gamma$, the mean number of photons illuminating the SiPM, is
 $N_\mathit{seed} = \mathit{pde} \cdot N_\gamma $, with \emph{pde}, the photon-detection efficiency of the SiPM without saturation effects.
 For the ratio $N_\mathit{meas} = Q_\mathit{meas}/Q_1$ also terms like \emph{effective number of fired cells}, \emph{ effective number of Geiger discharges} or \emph{detected number of photons}, are used.
 Frequently the ratio $N_\mathit{meas} / N_\mathit{pix}$ is shown in order to compare the saturation effects of SiPMs with different $N_\mathit{pix}$.

 The starting point of the phenomenological parameterisations is (Ref.\cite{Kotera:2016})
 \begin{equation} \label{equ:Nmeas0}
   N_\mathit{meas} = N_\mathit{pix} \cdot \big( 1 - e^{- N_\mathit{seed}/N_\mathit{pix}} \big),
 \end{equation}
 where it is assumed that the signal produced in a pixel is independent of the number of seed Geiger discharges in this pixel.
 Thus, the effects of pixel recovery after a Geiger discharge and of after-pulsing and cross talk are ignored.
 An improved version of this parametrisation is
 \begin{equation}\label{Nmeas1}
   N_\mathit{meas} = N_\mathit{pix}^\mathit{eff} \cdot \big( 1 - e^{- N_\mathit{seed}/N_\mathit{pix}^\mathit{eff}} \big),
 \end{equation}
 which is able to describe several measurements up to $N_\mathit{seed}/N_\mathit{pix} \approx 1.75$.
 As demonstrated in Ref.\,\cite{Kotera:2016}, the introduction of further phenomenological parameters allows describing experimental results up to $N_\mathit{seed}/N_\mathit{pix} \approx 10$.
 However, the effects of high dark-count rates are not considered, which ar relevant for the use of SiPMs at hadron colliders and in the presence of background light.

 In Ref.\,\cite{Rosado:2019} a parametrisation with three parameters is presented, which includes losses of gain and photo-detection efficiency, \emph{pde}, during the pixel recovery as well as the effects of correlated and uncorrelated noise.
 The three parameters are \emph{pde} at the nominal voltage, the saturation level of the SiPM response and the relative contribution of the correlated noise to the output charge.
 The parameters can either be obtained from a fit to the mean output charge as a function of the number of incident photons, or derived from charge spectra with resolved photo-electron peaks.
 The parametrisation is able to describe adequately the mean charge obtained for LYSO crystals exposed to different radioactive sources ($^{22}$Na, $^{60}$Co, $^{137}$Cs, and  $^{226}$Ra) read-out by a SiPM with 25\,$\upmu $m and 50\,$\upmu $m pixel size for a wide range of over-voltages.
 The LYSO decay time is 42\,ns and the maximum non-linearity (ratio of measured charge to expected charge without saturation effects) reached for the 50\,$\upmu$m SiPM was 0.45.
 A similar parametrisation was also to able to describe the saturation effects for the SiPMs illuminated by an LED operated in DC mode.

  A number of simulation programs are documented in the literature.
 An incomplete list follows.
 A Monte Carlo program to simulate the multiplication process which is responsible for Geiger discharges is presented in Ref.~\cite{Spinelli:1997}, and analytical calculations can be found in Ref.\,\cite{Windischhofer:2023}.
 Programs which simulate the shape of the  transients for different options for the readout electronics are discussed in Refs.~\cite{Acerbi:2019, Calo:2019}, and references therein.
 Monte Carlo programs addressing the readout of light from scintillators with SiPMs are documented in Refs.~\cite{Pulko:2012, Niggemann:2015}, where the last one has been implemented in the GEANT4 framework~\cite{Dietz:2017}.
 The simulation discussed in Ref.~\cite{Gundacker:2013} puts the main emphasis on the optimisation of the time resolution using SiPMs for PET scanners.

 In Refs.~\cite{vanDam:2010, Jha:2013} the most complete simulation so far is presented.
 It is based on a SPICE model for the pulse shape and, using Monte Carlo methods, Geiger discharges from photons and dark counts, after-pulses and prompt cross-talk are simulated in the individual SiPM pixels.
 The effects of pixel recharging, and the reduction of over-voltage in the pixels due to the voltage drop over the input resistance are taken into account in the program.
 In addition to the mean response for different readout schemes, also the variance of spectra and transients can be obtained.

 The comparison of the model to measurement with a LaBr$_3$:5\%Ce~calorimeter exposed to $\gamma$-rays with energies between 27.3 and 1836~keV from 7 radioactive sources is impressive~\cite{vanDam:2010}:
 For a SiPM with $N_\mathit{pix} = 3600$, the response for low $N_\gamma$ where $N_\mathit{meas}/N_\mathit{seed} \approx 1.2$ up to $N_\gamma = 1.3 \times 10^5$, where $N_\mathit{meas}/N_\mathit{seed} \approx 0.4$ is precisely described.
 The model has been successfully used to predict and optimise the response of calorimeters read out with different SiPMs for different deposited energies.
 Given all the details of the simulation, the program is quite CPU-time intensive.

 In this paper a method is described which allows simulating the response of SiPMs exposed to high photon intensities and high dark-count rates, which has many features of the simulation just described, but requires significantly less computing resources.
 As in these conditions the peaks of individual Geiger discharges can not be separated, the moments of the SiPM response are simulated.
 The first moment corresponds to the mean response.
 Its dependence on light intensity, which can be compared to experimental results, describes the non-linearity.
 Also the higher central moments can be compared to the experimental results, and, with the methods discussed in Ref.\cite{Vinogradov:2023}, it can be investigated if other SiPM parameters like photon intensity, gain and correlated noise can be extracted from the experimental data.
 The program also allows to investigate the photon-signal reduction as a function of the dark-count rate which increases rapidly with radiation damage.
 In addition, the simulation program can be used to find phenomenological parameterisations of the SiPM non-linearity.
 Last but not least, with the simulation program the readout and SiPM operating conditions for a given application can be optimised.

 The paper is structured in the following way:
 In the next section the program and the parameters used are described, and the method is illustrated by comparisons to published results for SiPMs exposed to photon pulses with two significantly different pulse shapes and with intensities varying by several orders of magnitude.
 This is followed by comparisons of simulations with further literature data at high light intensities for SiPMs with different number of pixels and pixel sizes, different light-pulse shapes and different over-voltages.
 Finally, the main results are summarized.

 \section{The SiPM response model}
  \label{sect:Model}

  \subsection{Overview}
  \label{subsect:ModelOverview}

 Before entering into details, an overview over the simulation is presented.
  For the symbols see Table\,\ref{tab:Parameters}.
 The program simulates the response of SiPMs with $N_\mathit{pix}$ pixels for photons from a pulsed light source and the dark-count rate \emph{DCR}.
  Its intended application is the study of SiPM-saturation effects for high photon intensities at low and also at high dark-count rates.
 As under these conditions individual Geiger discharges cannot be separated, the moments of the response of the combined system SiPM and readout electronics are simulated.
  The mean number of primary Geiger discharges from the photons in the absence of saturation effects is denoted $N_\mathit{seed}$.
 The first moment divided by $N_\mathit{seed}$, is the response non-linearity.
  Higher moments can be used to investigate if $N_\mathit{seed}$ and cross-talk can be estimated from data as discussed in Ref.\,\cite{Vinogradov:2023}.

 First, the response of a single pixel for one measurement, called \emph{event} in the following, is simulated.
  For one pixel the mean number of potential primary Geiger discharges from light is $N_\mathit{seed}/N_\mathit{pix}$, and the dark-count rate $\mathit{DCR}/N_\mathit{pix}$.
 From the distribution of the response for many events, the \emph{single-pixel response moments} are obtained.
 From the single-pixel moments, the moments of the response of the entire SiPM are calculated.
 In Ref.\,\cite{Laury:1976} it is shown that for the first three moments the convolution of two distributions are the sums of the moments of the two distributions.
 Thus, assuming that all pixels behave the same, the first three moments of the response of the entire SiPM are simply $N_\mathit{pix}$ times the single-pixel moments.
 \\

 The different simulation steps are:

 \begin{enumerate}
 \item
  The single-pixel response for one event is simulated by first generating the times of primary Geiger-discharge candidates from photons and dark counts, and for every potential primary Geiger discharge, the times of the after-pulse candidates.
 From the time-ordered list the \emph{Geiger Array} is calculated, which contains for every entry with index $i$, the charge $A_i$ of the Geiger discharge in units of the gain at nominal over-voltage, $\mathit{OV}_0$, its time $t_i$, the mean probability for it occurring, $\mathit{prob}_i$, and  for after-pulses the link to the corresponding primary discharge.
  When calculating $\mathit{prob}_i$, the reduction in Geiger-discharge probability due  to the recharging of the pixel after a preceding Geiger discharge is taken into account.
 For after-pulses, in addition the charge of the primary Geiger-discharge and the after-pulsing probability parameter, $p_\mathit{Ap}$, and the after-pulsing time constant, $\tau_\mathit{Ap}$, enter in the calculation of $\mathit{prob}_i$.
  Next, a random number, which is uniformly distributed between zero and one, is generated.
 If this number exceeds $\mathit{prob}_i$, $A_i$ is set to zero, otherwise $A_i$ is calculated taking into account the reduction of the charge due to the recharging of the pixel.
 The concept of the Geiger Array has been introduced in Ref.\,\cite{Garutti:2020}, but has been used effectively already in Ref.\,\cite{Jha:2013}.
 \item
 From the Geiger Array and the pulse shape of a single Geiger discharge, the time dependence of the over-voltage of a single pixel for a single event is calculated.
 From the average of many events the time dependence of the over-voltage for the entire SiPM, $\mathit{OV}(t)$, and the voltage drop over the input impedance of the readout, $V_L(t)$, are obtained.
 \item
 Next, step 1. is executed again, but this time, the reduction of $A_i$ and of $\mathit{prob}_i$ due to $V_L(t)$, which reduces the over-voltage, is taken into account.
 \end{enumerate}

 So far, cross-talk has not been taken into account.
 As shown in Ref.\,\cite{Jha:2013}, the detailed simulation of cross-talk is complex and CPU-time consuming, as Geiger discharges in all pixels have to be simulated and discharges in one pixel influence the amplitudes and probabilities of Geiger discharges in other pixels.
 Already accepted discharges have to be removed, and already rejected ones resurrected.
 Therefore, a simplified approach is taken, which only considers the average reduction of the Geiger-discharge probability and of the signal charge due to the reduced $\mathit{OV}(t)$ when evaluating the cross-talk.

 There are two types of cross-talk:
 (1) from the single pixel under consideration to all other pixels, and
 (2) from all other pixels to the pixel under consideration.

 \begin{enumerate}
  \setcounter{enumi}{3}
 \item
  For simulating the cross-talk from all other pixels to the single pixel under consideration, an additional Geiger Array is generated as described in 3.
  From this Geiger Array, which is used to calculate the cross-talk from all other pixels, the times and charges of potential prompt and delayed cross-talk discharges are obtained.
   The mean probability that a prompt cross-talk occurs is given by $p_\mathit{pXT}$ times the normalised charge of the Geiger discharge causing the prompt cross-talk times the reduction in probability due to the reduced over-voltage, $\mathit{OV}(t_\mathit{pXT})$, calculated in 2.
  For the delayed cross-talk $p_\mathit{pXT}$ is replaced by $p_\mathit{dXT}$, and $t_\mathit{pXT}$ by $t_\mathit{dXT}$.
 The time of the prompt cross-talk, $t_\mathit{pXT}$, is the time of the Geiger discharge causing the prompt cross-talk plus the discharge built-up time, which is assumed to  be 100\,ps.
 The time of the delayed cross-talk, $t_\mathit{dXT}$, is obtained by adding to the time of the Geiger discharge which causes the delayed cross-talk, the discharge built-up time and a random number distributed according to $ e^{- (t - t_\mathit{pXT})/\tau _\mathit{dXT}} \cdot \Theta(t - t_\mathit{pXT})$, where $\tau _\mathit{dXT}$ is the time constant of the delayed cross-talk and $\Theta(t)$ the Heaviside step function.
 The cross-talk Geiger Array is obtained in a way similar to what is described in 1.
 A random number uniformly distributed between zero and one is generated, and $A_i$ is set to zero if the random number exceeds the mean probability. Otherwise $A_i = \mathit{OV}(t_i) / \mathit{OV}_0$, where $\mathit{OV}_0$ is the over-voltage in the absence of Geiger discharges.

  In a similar way, the Geiger Array for the cross-talk from the pixel under consideration to the other pixels is calculated.
  Both cross-talk Geiger Arrays are added to the Geiger Array from 3.

 \item
  From the complete Geiger Array the single-pixel response for one event can be calculated.
  As discussed at the beginning of this section, by generating many events, the moments of the single-pixel response of the combined system SiPM and readout is obtained.

  What is understood under \emph{response}, depends on the experimental set-up simulated.
   If the SiPM transient is integrated in a gate, the single-pixel transient $I(t) = \Sigma A_i \cdot f(t-t_i) \cdot \Theta(t-t_i)$ is integrated in the gate.
    Here $f(t)\cdot \Theta(t)$ is the normalised pulse shape for a Geiger discharge at $t = 0$.
  This is presently the most common way of recording SiPM signals.
   If the signal is shaped and the maximum of the amplitude is recorded, the transient $I(t)$ has to be shaped and the maximum determined.
 \end{enumerate}

  It is noted that single-pixel Geiger Arrays can also be used to simulate spectra for the entire SiPM, which can be useful for low light intensities and low dark-count rates.
  For the simulation of a single event, the number of seeds in the individual pixels is obtained by generating $N_\mathit{pix}$ random numbers from a Poisson distribution with mean $N_\mathit{seed} / N_\mathit{pix}$, and for the dark counts with the mean $(DCR/N_\mathit{pix}) \cdot \Delta t$, where $\Delta t$ is the time interval simulated.
  Next, $N_\mathit{pix}$ single-pixel Geiger Arrays are simulated, summed and convolved with the electronics noise, to obtain the Geiger Array for the entire SiPM.
  From this Geiger Array the response for a single event is calculated as described above, and the response spectrum is obtained from many events.
  For low values of $N_\mathit{seed}$ and \emph{DCR} most of the pixels will have no Geiger discharge and the simulation will be fast, whereas for high values most pixels will have discharges and the simulation will be CPU intensive.

  It is stressed that for all simulations it is assumed that the response for all pixels is the same.
  This means that the photon distribution is uniform over the SiPM and that all pixels have the same dark-count rate.
  In addition, edge effects, which are relevant for SiPMs with small number of pixels, are ignored.


  \subsection{Details of the model}
  \label{subsect:ModelDetails}

 To characterise the response of a SiPM, read out by a given electronics and exposed to given light pulses, requires many parameters.
 References~\cite{Klanner:2019, Acerbi:2019, Calo:2019, Piemonte:2019} give many details of the functioning of SiPMs, and discuss methods how to extract the parameter values from experimental data.
 Table~\ref{tab:Parameters} summarises the parameters used.

 \begin{table}[!ht]
 \caption{Parameters used in the simulation program, and values for simulating the data of Ref.\,\cite{Kotera:2016}.}
  \centering
   \begin{tabular}{c|c|r}
  Symbol & Description & Value\hspace{3mm} \\
    \hline \hline
  $N_\mathit{pix}$ & Number of pixels & 1600 \\
  $\mathit{OV_0}$ & Nominal over-voltage & 3 V \\
  \emph{DCR} & Dark-count rate & 100 kHz \\
  $\sigma _0$ & Electronics noise & 0.075 gain \\
  $R_L$ & Load resistance & 50 $\Omega$ \\
 \hline
  $R_q$ & Quenching resistor & 670 k$\Omega$ \\
  $\tau_s$ & Time const. slow component & 17.5 ns \\
  $\tau_f$ & Time const. fast component & --\hspace{4mm}  \\
  $r_f$ & Contribution fast component & 0 \\
  $t_\mathit{GD}$ & Geiger discharge build-up time & 0.1\,ns \\
 \hline
  $-t_0$ & Start time simulation & 100 ns \\
  $t_\mathit{start}$ & Start-time gate & 0 ns \\
  $t_\mathit{gate}$ & Gate length & 150 ns \\
 \hline
  $N_\mathit{seed}$ & No. seed pixels photons & 10 to $10^{\,5}$ \\
  $t_L$ & Start-time light pulse & 0.5 ns \\
  $\tau_\mathit{SCI}$ & Time const. scintillator & 2.2 ns \\
  $\tau_\mathit{WLS}$ & Time const. wave-shifter & 11 ns \\
 \hline
  $\tau _\mathit{rec}$ & Time const. pulse recovery & 15 ns \\
  $p_\mathit{Ap} $ & After-pulsing probability & 0.05 \\
  $\tau_\mathit{Ap} $ & Time const. after-pulsing & 15 ns \\
  $p_\mathit{pXT} $ & Prompt-XT probability & 0.02 \\
  $p_\mathit{dXT} $ & Delayed-XT probability & 0.02 \\
  $\tau_\mathit{dXT} $ & Time const. delayed XT & 25 ns \\
 \hline
  \end{tabular}
  \label{tab:Parameters}
 \end{table}

\begin{figure}[!ht]
   \centering
    \includegraphics[width=0.6\textwidth]{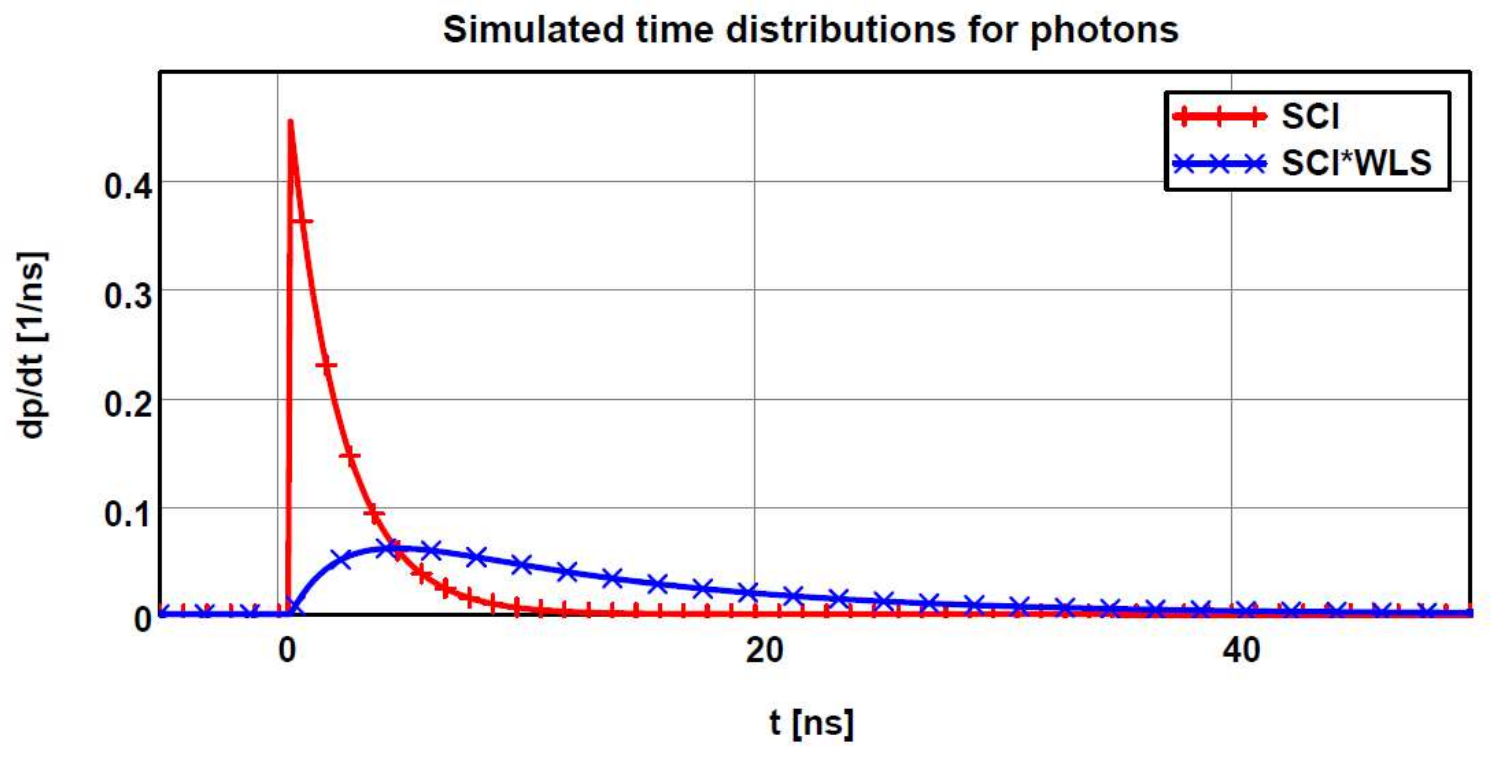}
   \caption{Normalised time distributions of the number of photons for the experiments of Ref.\,\cite{Kotera:2016} for the conditions \emph{SCI} and \emph{SCI*WLS}.
   The start of the simulated photon pulse is at $t = 0.5$\,ns.}
  \label{fig:Ngamma-t}
 \end{figure}

 The individual simulation steps are described in detail in the following and illustrated using the measurements and experimental results of Ref.\,\cite{Kotera:2016}, where SiPMs corresponding to the Model S10362-11-25P, fabricated by Hamamatsu in 2008, have been investigated.
 They have $N_\mathit{pix} = 1600$ pixels of $25\,\upmu$m pitch, and a sensitive area of 1\,mm$^2$.
 The values of the SiPM parameters used in the simulation are given in Table~\ref{tab:Parameters}.
 Two light sources have been employed:
 A $3 \times 10 \times 45$\,mm$^3$ SCSN-38 scintillator excited by laser light with a wavelength of 408\,nm and a duration with a FWHM of 31\,ps with,
 (1) the SiPM directly coupled to the scintillator, called \emph{SCI} in the following, and
 (2) the SiPM coupled to the scintillator via a Kuraray Y-11 wavelength-shifting fiber, called \emph{SCI*WLS}.
 For simulating the photon pulses the values of the time constants $\tau _\mathit{SCI} = $\,2.2\,ns and $\tau _\mathit{WLS} =$\,11\,ns are used.
 Figure\,\ref{fig:Ngamma-t} shows the expected normalised time distributions of the lighte pulses:
 d$p_\mathit{SCI} /  \mathrm{d} t = e^{-t/\tau _\mathit{SCI}}/ \tau _\mathit{SCI}$ for \emph{SCI}, and
 d$p_\mathit{SCI*WLS} / \mathrm{d} t = (e^{-t/\tau _\mathit{SCI}} -  e^{-t/\tau _\mathit{WLS}}) / (\tau _\mathit{SCI} - \tau _\mathit{WLS}) $ for \emph{SCI*WLS}.
 It is noted that the maximal photon flux for \emph{SCI} is about an order of magnitude larger than for \emph{SCI*WLS}, and that the ratio of the time spread of the pulses
 $\sigma _\mathit{SCI*WLS} / \sigma _\mathit{SCI} = \sqrt{1 + (\tau _\mathit{WLS} / \tau _\mathit{SCI})^2} = 5.15$.
 Thus, significant differences of the non-linearity for the two conditions are expected.

 The laser-light intensity was changed over a wide range by the relative angle of two polarisation filters and was monitored by a vacuum photo-multiplier with excellent linearity over the range of the measurements.
 The SiPM was operated at an over-voltage of 3\,V and the SiPM pulse integrated over 150\,ns using a charge sensitive ADC, where an input resistance of $50\,\Omega$ is assumed in the simulation.
 For determining the charge for a single Geiger discharge, a charge-sensitive amplifier had to  be used.
 The ratio of the measured charge to the charge from a single Geiger discharge is called $N_\mathit{fired}$, the \emph{"number of fired pixels"}, which is a misnomer:
 There are many more pixels with Geiger discharges than \emph{fired pixels}, However, the charge they generate is reduced because of the recharging of the pixel.

 \begin{figure}[!ht]
   \centering
   \begin{subfigure}[a]{0.5\textwidth}
    \includegraphics[width=\textwidth]{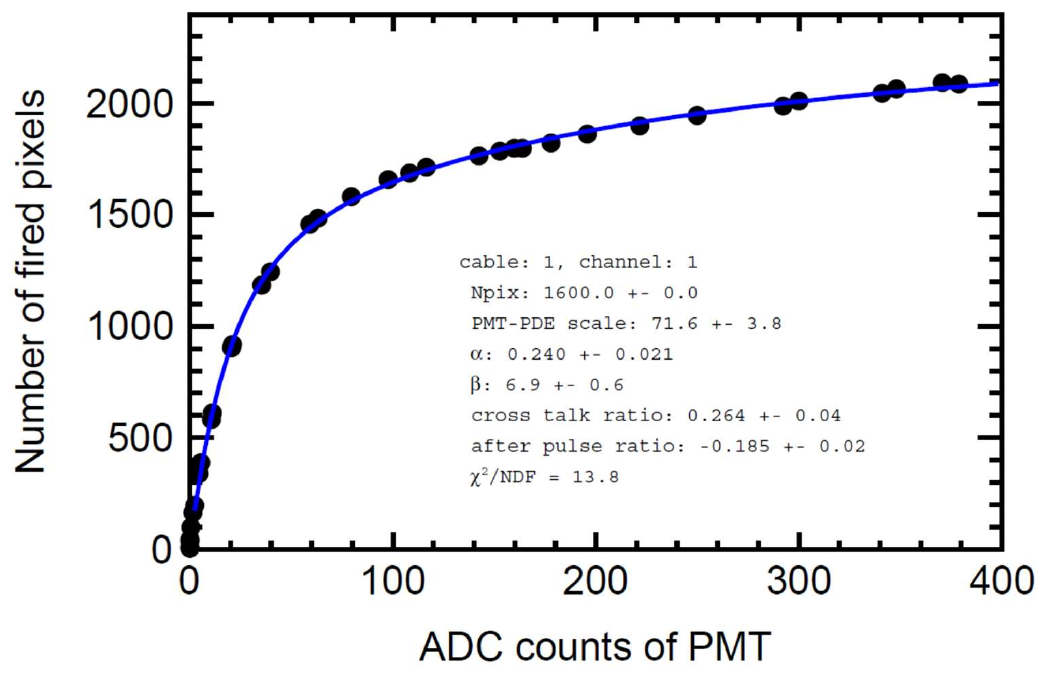}
    \caption{ }
    \label{fig:KoteraSCI}
   \end{subfigure}%
    ~
   \begin{subfigure}[a]{0.5\textwidth}
    \includegraphics[width=\textwidth]{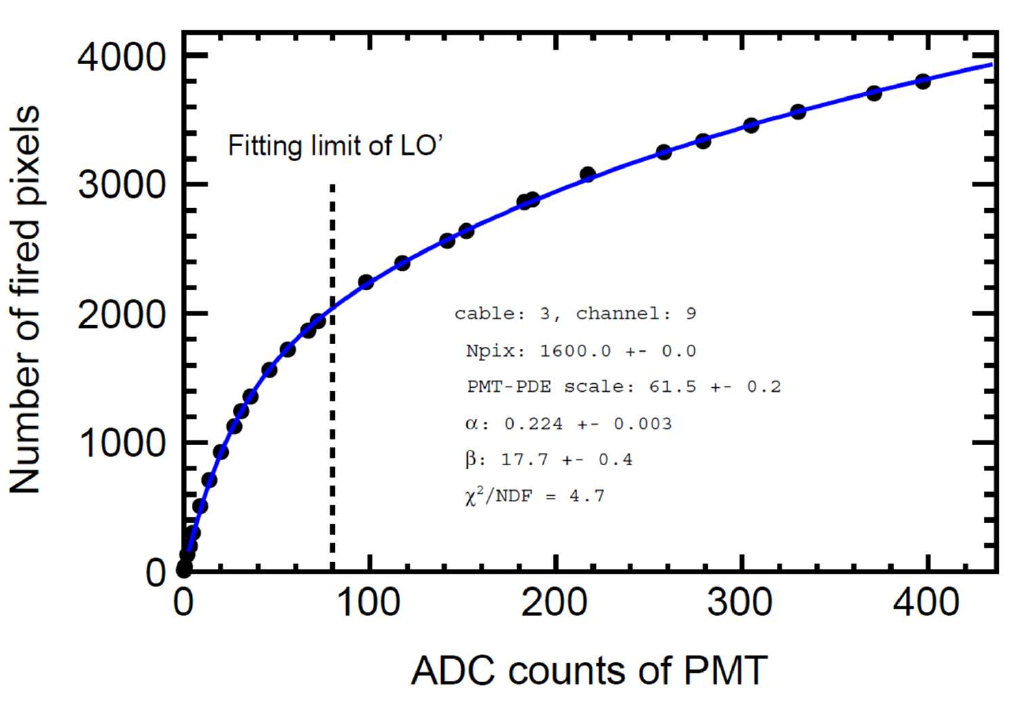}
    \caption{ }
    \label{fig:KoteraWLS}
   \end{subfigure}%
   \caption{Results for the mean charge measured by the SiPM in units of the charge of a single Geiger discharge from Ref.\,\cite{Kotera:2016} as a function of the photon intensity for the conditions described in the text for
   (a) \emph{SCI}, and (b) \emph{SCI*WLS}.
   The $x$\,axes are the charge measured by the vacuum photo-multiplier, which is proportional to the number of photons hitting the scintillator.
   The filled circles are the data points and the curves are the results of phenomenological parameterisations fitted to the data, which are used for the comparison with the simulations of this paper.
   Note, that the vertical scales differ by about a factor 2.
   The figure is taken from the arXiv\,version of Ref.\,\cite{Kotera:2016}.
   }
  \label{fig:Kotera}
 \end{figure}

 Figure\,\ref{fig:Kotera} shows $N_\mathit{fired}$ for \emph{SCI} and \emph{SCI*WLS} as a function of the charge from the vacuum photo-multiplier.
 In addition, phenomenological fits can be seen which provide a good description of the data.
 As expected, the non-linearity for the two conditions is very different.
 For both cases $N_\mathit{fired}$ reaches values well above the number of pixels, $N_\mathit{pix} = 1600$.

 For simulating a Geiger discharge at $t=0$, the following parametrisation of the SiPM pulse is used in the program:
 \begin{equation}\label{eq:SiPMpuls}
   I(t) = \left(\frac{1-r_f}{\tau_s}\cdot e^{-t/\tau_s} + \frac{r_f}{\tau_f} \cdot e^{-t/\tau_f} \right) \cdot \Theta(t),
 \end{equation}
 which assumes that the SiPM gain is one, or in other words, the results are given in units of SiPM gain at the nominal over-voltage.
 The contribution of a fast component is $r_f$, and $\tau _s$ and $\tau _f$ are the time constants of the slow and of the fast component, respectively.
 For the parameter values used in the simulation see Table\,\ref{tab:Parameters}.
 The finite rise-time, which according to Refs.\,\cite{Windischhofer:2023, Jha:2013} is of order 0.1\,ns, is neglected in this parametrisation.

 The simulation is performed in the time interval $-t_0 \leq t \leq t_\mathit{start} + t_\mathit{gate}$, and the light pulse starts at $t = t_L$.
 The time interval between  $- t_0$ and the start of the gate, $t_\mathit{start}$, has to be chosen of sufficient length so that the signals from Geiger discharges for $t \leq -t_0$ can be neglected.
 For specifying the end of the time interval, $t_\mathit{end} = t_\mathit{start} + t_\mathit{gate}$, the length of the gate, $t_\mathit{gate}$, is used, as frequently the SiPM current integrated in a gate is recorded.
 However, the method can also be used for recording amplitudes after pulse shaping.

 Next, the times of potential Geiger discharges from photons, dark counts and after-pulses in a single pixel are simulated.
 The number of dark counts is sampled from a Poisson distribution with the mean $(t_0 + t_\mathit{end}) \cdot \mathit{DCR}/N_\mathit{pix}$, and the times from random numbers uniformly distributed between $-t_0$ and  $t_\mathit{end}$.
 The number of potential Geiger discharge from photons is  sampled from a Poisson distribution with the mean $N_\mathit{seed}/N_\mathit{pix}$, where $N_\mathit{seed}$ is the photon-induced number of Geiger discharges in the absence of saturation effects.
 Other distributions, like the energy loss distribution of charged particles in a scintillator, can easily be implemented.
 The time distribution of the photon-induced seeds is randomly sampled from the expected time distribution of the photon pulse. As examples, the two distributions for simulating the experiment of Ref.\,\cite{Kotera:2016} are shown in Fig.\,\ref{fig:Ngamma-t}.

 The times of potential after-pulses are simulated by adding to the times of the primary Geiger discharges the time $t_\mathit{Ap}$ randomly sampled from the distribution

 \begin{equation}\label{eq:tAp}
   \frac{\mathrm{d}p}{\mathrm{d}t_\mathit{Ap}} =\left( 1 - e^{-t _\mathit{Ap}/ \tau _\mathit{rec} }\right) \cdot e^{-t _\mathit{Ap}/ \tau _\mathit{Ap}} \cdot \left(\tau_\mathit{Ap} + \tau_\mathit{rec}\right) / \tau_\mathit{Ap}^2.
 \end{equation}

 This parametrisation assumes that after-pulses are caused by charges trapped by states in the silicon band-gap during the primary Geiger discharge and de-trapped with the single time constant $\tau _\mathit{Ap}$.
 The term $1 - e^{-t _\mathit{Ap}/ \tau_ \mathit{rec}}$ describes the reduction in discharge probability because of the recharging of the pixel.
 In Ref.\,\cite{Rolph:2023} it is shown how $\tau_ \mathit{rec}$ can be measured, and that $\tau_ \mathit{rec} \approx f_\mathit{rec} \cdot \tau_ s $.
 The value of $f_\mathit{rec}$ depends on the design of the SiPM and decreases with increasing over-voltage.
 A typical value is 0.65 for an over-voltage of 3\,V, which is used in the simulation.

 In the following it is described, how, from the time-ordered list of the times of potential Geiger discharges, the Geiger Array, introduced in section\,\ref{subsect:ModelOverview}, is obtained.
 The simulation takes into account the reduction of the charge and of the probability by the recharging of the pixel and the after-pulsing-probability parameter $p_\mathit{Ap}$.
 The Geiger Array contains for every entry $i$ in the time-ordered list of potential Geiger discharges,
  the Geiger-discharge charge in units of SiPM gain at nominal over-voltage, $A_i$,
  its time, $t _i$,
  the link to the corresponding primary Geiger discharge for after-pulses ,
  and $\mathit{prob} _i$, the mean probability for the occurrence of the Geiger discharge.

  For the first entry, which is either a Geiger discharge from a photon or a dark count, the charge and the probability are set to $A_0 = 1 - e^{- N_\mathit{pix} / (DCR \cdot \tau_s)}$ and to $\mathit{prob} _0 = 1$, respectively.
  This choice for $A_0$, where $N_\mathit{pix} / \mathit{DCR}$ is the mean time between dark counts, improves the transient behaviour for high dark-count rates at the times following $-t_0$.
  For the following entries, $i$, the time difference, $\Delta t$, to the $j$-th entry in the Geiger Array preceding $i$ with $A_j > 0$ is calculated.
  If this entry corresponds to a primary Geiger discharge, $\mathit{prob} _i = 1 - e^{- \Delta t / \tau _\mathit{rec}} $.
  If a uniformly generated random between zero and 1 exceeds $\mathit{prob} _i$, $A_i = 0 $, otherwise $A_i = 1 - e^{-\Delta t / \tau _s}$.
  If the $i$-th entry corresponds to an after-pulse and the corresponding primary Geiger discharge has $A_k = 0$, the values of the Geiger array are $A_i = 0$ and $\mathit{prob} _i = 0$,
  otherwise $\mathit{prob} _i = p_{Ap} \cdot A_k \cdot (1 - e^{- \Delta t /\tau _\mathit{rec}})$.
  If a uniformly generated random between zero and 1 exceeds $\mathit{prob} _i$, $A_i = 0 $, otherwise $A_i = 1 - e^{-\Delta t / \tau _s}$.
 At this step gain fluctuations could be implemented by multiplying $A_i$ by a Gaussian random number with mean one and $\sigma = \sigma_1$, where $\sigma_1$ is the gain spread.
 However, the effect on the final results is negligible and therefore not done.

\begin{figure}[!ht]
   \centering
   \begin{subfigure}[a]{0.5\textwidth}
    \includegraphics[width=\textwidth]{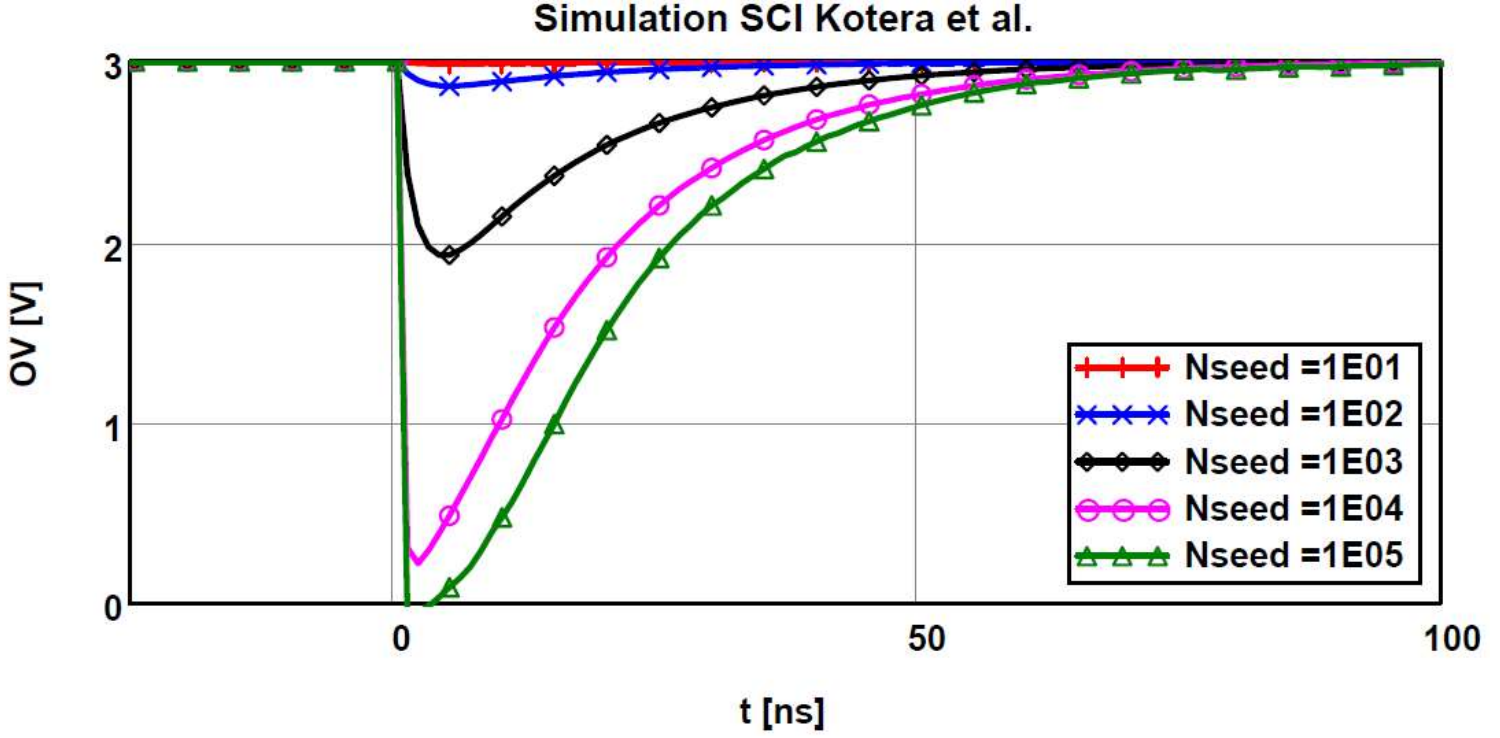}
    \caption{ }
    \label{fig:OV-SCI}
   \end{subfigure}%
    ~
   \begin{subfigure}[a]{0.5\textwidth}
    \includegraphics[width=\textwidth]{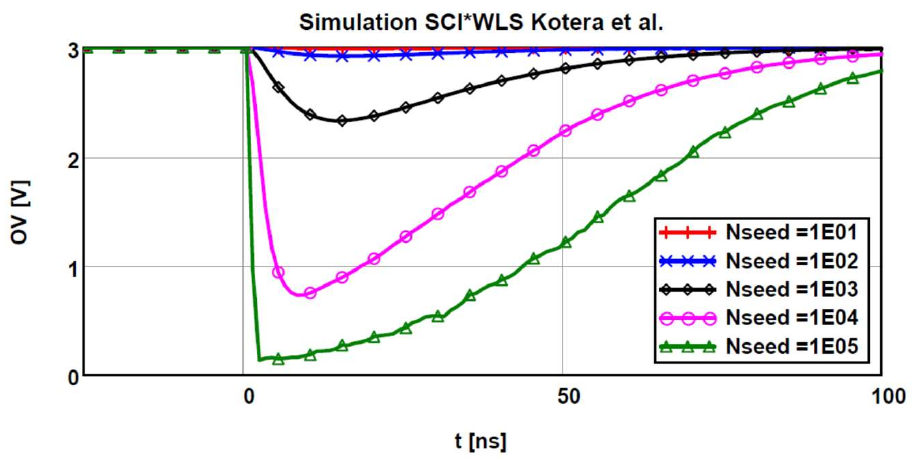}
    \caption{ }
    \label{fig:OV-WLS}
   \end{subfigure}%
   \caption{Time dependence of the mean over-voltage for different $N_\mathit{seed}$ values for the conditions (a) \emph{SCI} and (b) \emph{SCI*WLS}.
   }
  \label{fig:OV}
 \end{figure}

 Next the time dependence of the mean over-voltage for the entire SiPM, $\mathit{OV}(t)$, and the voltage drop over the load resistance of the readout, $V_L(t)$, are calculated.
 For a single Geiger discharge at $t = 0$ the time-dependent over-voltage is $\mathit{OV}_0 \cdot \left(1 - e^{-t/\tau_s} \right) \cdot \Theta(t)$, where $\mathit{OV}_0$ is the nominal over-voltage.
 For several Geiger discharges the time-dependent over-voltage of a single pixel for one event can be obtained from the Geiger Array:
 \begin{equation}\label{eq:OV-t}
   \mathit{OV}_1(t) = \mathit{OV}_0 \cdot \sum_i \left( A_i  \cdot \left( 1 - e^{-(t - t_i)/\tau_s}\right) \cdot \Theta (t - t_i) \right).
 \end{equation}
 Assuming that all pixels behave identically, the averaging over many events gives $\mathit{OV}(t)$.
 Figure\,\ref{fig:OV} shows for \emph{SCI} and \emph{SCI*WLS} the simulated $\mathit{OV}(t)$  for different values of $N_\mathit{seed}$, the number of photon-induced Geiger discharges in the absence of saturation.
 As expected, the decrease in over-voltage increases faster with $N_\mathit{seed}$ for the shorter \emph{SCI} pulse, and the duration of the reduction is longer for the slower \emph{SCI*WLS} pulse.
 At a rate of seed pulses per pixel and nanosecond of about five, which is reached at $N_\mathit{seed}$ values of $10^4$ for \emph{SCI} and $10^5$ for \emph{SCI*WLS}, the over-voltage drops to zero and the SiPM is saturated.

\begin{figure}[!ht]
   \centering
   \begin{subfigure}[a]{0.5\textwidth}
    \includegraphics[width=\textwidth]{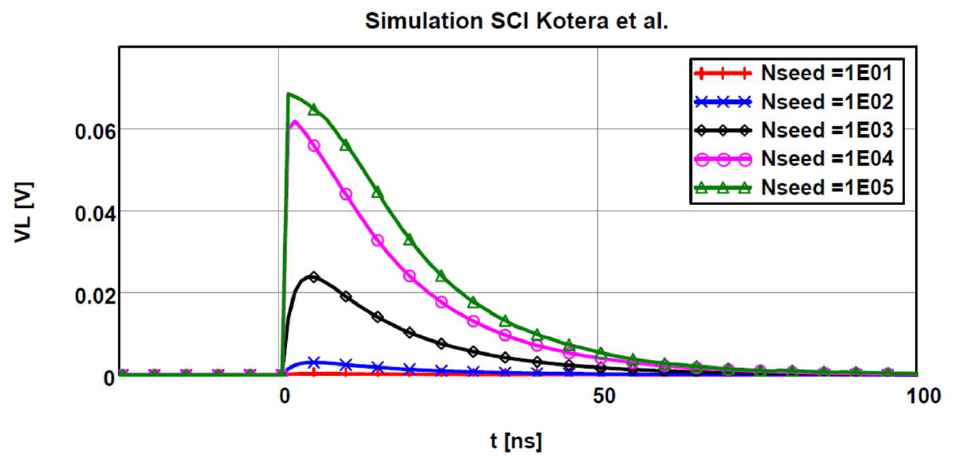}
    \caption{ }
    \label{fig:VL-SCI}
   \end{subfigure}%
    ~
   \begin{subfigure}[a]{0.5\textwidth}
    \includegraphics[width=\textwidth]{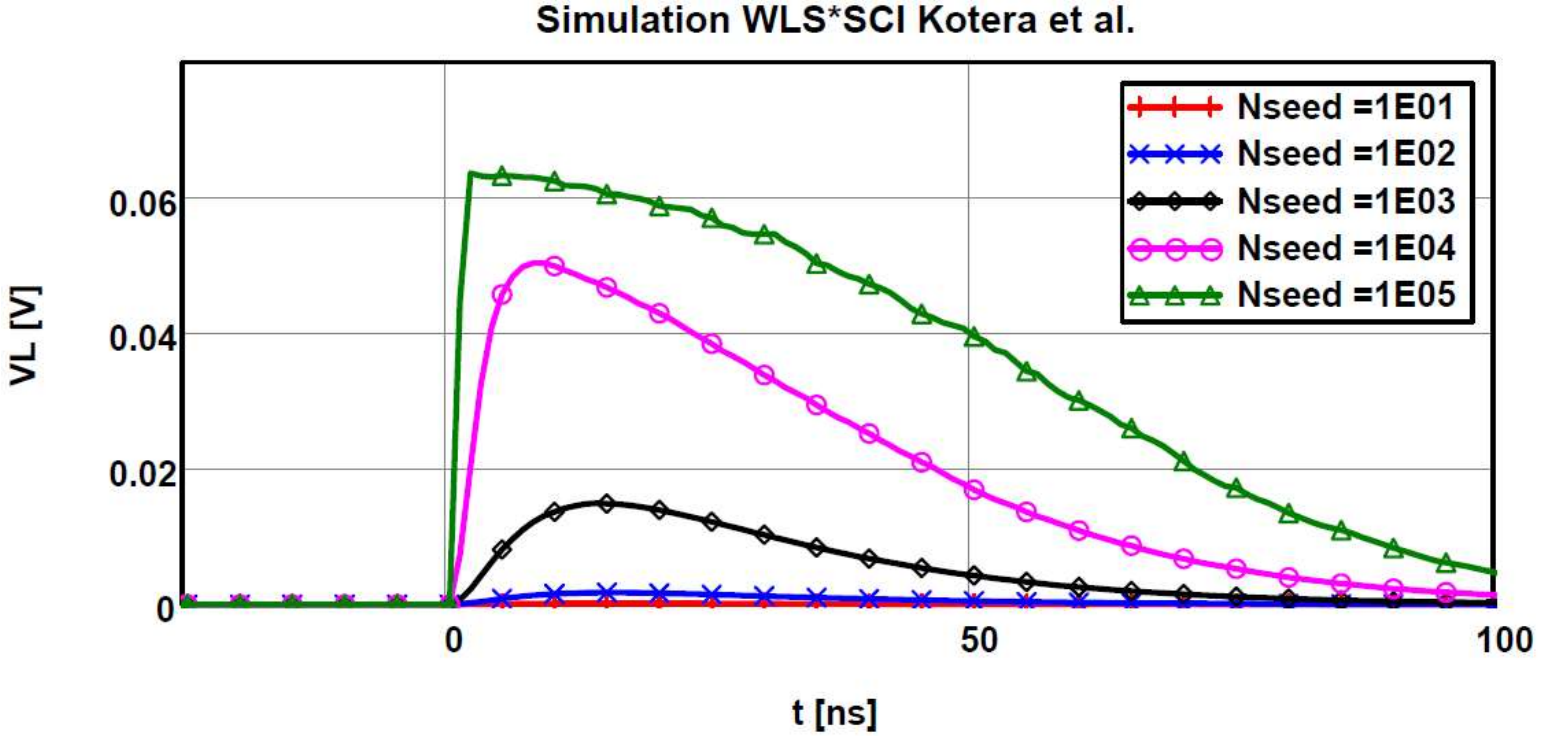}
    \caption{ }
    \label{fig:VL-WLS}
   \end{subfigure}%
   \caption{Time dependence of the mean voltage drop over the $50\,\Omega$ input resistance of the readout electronics for different $N_\mathit{seed}$ values for the conditions (a) \emph{SCI} and (b) \emph{SCI*WLS}.
   }
  \label{fig:VL}
 \end{figure}

 From $V_\mathit{bias} = V_L(t) + V_\mathit{bd} + \mathit{OV}(t) + V_q(t)$  and $V_\mathit{bias} = V_\mathit{bd} + OV_0$ follows:

 \begin{equation}\label{eq:VL}
   V_L (t) = \frac{OV_0 - OV(t)}{1 + R_Q / R_L},
 \end{equation}
 where $V_\mathit{bias}$ is the biasing voltage, $V_L$ the voltage drop over the load resistance $R_L$, $V_\mathit{bd}$ the break-down voltage, and $V_q(t)$ the time-dependent voltage over the quenching resistor $R_q$.
 The parallel resistance of the $N_\mathit{pix}$ quenching resistors for the entire SiPM is denoted $R_Q = R_q/N_\mathit{pix}$.
 Using the value $R_q = 670$\,k$\Omega $ from Table \ref{tab:Parameters}, the time dependencies of $V_L(t)$ for \emph{SCI} and \emph{SCI*WLS} are obtained and shown in Fig.\,\ref{fig:VL}.
 It is noted that for a nominal over-voltage $\mathit{OV}_0 = 3$\,V, the value $V_L = 30$\,mV corresponds to a decrease in gain of 1\,\%.
 Although the effect is small for the SiPM considered, it can be large for SiPMs with many more pixels.
 For this reason it is implemented in the simulation.
 In order to take into account the effect of $V_L(t)$, the simulation of the Geiger Array is repeated, starting by generating the times for dark counts, photon-induced Geiger discharges and after-pulses, with $\mathit{OV_0}$ replaced by $\mathit{OV}_0 - V_L(t)$ when calculating the charges and probabilities of the Geiger Array.
 As shown in Fig.\,36 of Ref.\,\cite{Rolph:2023}, $\tau _\mathit{rec}$ increases with decreasing over-voltage.
 However, this effect is negligibly small and has been ignored when calculating the probabilities, and a constant $\tau _\mathit{rec}$ has been assumed.

 So far, cross-talk has not been taken into account.
 There are four types of cross-talk:
 from the pixel under consideration to the remaining pixels and from the remaining pixels to the pixel under consideration, and prompt and delayed cross-talk for each.

 Prompt cross-talk is caused by photons produced in the Geiger discharge which convert in the amplification region of a different pixel and cause a Geiger discharge there.
 Given a light path of typically only tens of micrometers, they can be considered  prompt.
 The light path can be directly through the silicon bulk, via reflection on the back or the front surface of the SiPM, or via a detector coupled to the SiPM.
 The introduction of trenches filled with light-absorbing material has significantly reduced the probability of prompt cross-talk~\cite{Acerbi:2019}.

 The dominant source of delayed cross-talk are photons from the primary Geiger discharge which convert to  electron-hole pairs in the non-depleted silicon.
 If minority charge carriers diffuse into the amplification region, they can produce a delayed Geiger discharge there.
 Delayed cross-talk in the same pixel as the primary Geiger discharge cannot be distinguished from after-pulses and is considered there.

 The detailed simulation of cross talk requires the simulation of all pixels with Geiger-discharge candidates and the complex removal and resurrection of Geiger discharges and changes of their amplitudes because of the cross-talk.
 In Ref.\,\cite{Jha:2013} it is shown that, even if only the neighbouring pixels are taken into account, this is quite complex and CPU-time intensive.
 Here a significantly simpler method is proposed.

 All remaining pixels are lumped into a single pixel, whose Geiger Array is simulated.
 For simulating the prompt crosstalk of the $i$-th element of this Geiger Array to the pixel under consideration, a binomial probability distribution with the parameter
 \begin{equation}\label{eq:pOV}
  \mathit{prob} \approx p_\mathit{pXT} \cdot A_i \cdot \frac{ 1 - e^{-\left(\mathit{OV}(t_\mathit{pXT}) - V_L(t_\mathit{pXT})\right) / V_0}} {1 - e^{-\mathit{OV}_0 / V_0}}
 \end{equation}
is assumed, and for the charge
 \begin{equation}\label{eq:AXT}
 A \approx \frac{\mathit{OV}(t_\mathit{pXT})- V_L(t_\mathit{pXT})}{\mathit{OV}_0}
 \end{equation}

 The individual terms of Eq.\,\ref{eq:pOV} are motivated in the following way:
 The probability of a prompt Geiger discharge to occur for the SiPM at the nominal over-voltage is $p_\mathrm{XT}$.
 The discharge probability is proportional to the number of charge carriers produced in the avalanche, which in units of nominal gain, is given by $A_i$.
 The probability also depends on $\mathit{OV}(t_\mathit{pXT})-V_L(t_\mathit{pXT})$, the effective over-voltage of the pixel under consideration.
 In Ref.\,\cite{Rolph:2023} it is observed that the voltage dependence of the photon-detection efficiency, which is proportional to the breakdown probability, can be described by $(1 - e^{-\mathit{OV}/V_0})$, where the value $V_0 = 2.91$\,V for a Hamamatsu MPPC S13360-1325PE has been determined.
 Thus, the breakdown probability at the effective over-voltage relative to its value at $\mathit{OV}_0$ is given by the third term of Eq.\,\ref{eq:pOV}.
 For lack of better knowledge, the value 2.91\,V for $V_0$ is used in the simulation.
 The amplitude of the cross-talk discharge is given by $\mathit{OV}_0$ minus the voltage drop over $R_L$, divided by the over-voltage in the absence of Geiger discharges, as shown by Eq.\,\ref{eq:AXT}.
 The treatment of cross-talk is very much simplified by using the average over-voltage reduction of the entire SiPM.
 Also the use of a binomial instead of a Borel distribution is a simplification, which however hardly affects the results, in particular for small $p_\mathit{pXT}$, high photon fluxes and high dark-count rates.

 For the delayed crosstalk from the $i$-th element of the Geiger Array of the simulated remaining pixel to the pixel under consideration, to every entry in the Geiger Array the delayed cross-talk time $t_\mathit{dXT}$ is generated by adding to $t_i$ a random number distributed according to $e^{-x/\tau_\mathit{dXT}} / \tau_\mathit{dXT}$.
 The parametrisation of the delay by an exponential is certainly a crude approximation, however, the topic has not yet been investigated so far.
 Similar to Eq.\,\ref{eq:pOV} the probability is calculated using

 \begin{equation}\label{eq:pdOV}
  \mathit{prob} \approx p_\mathit{dXT} \cdot A_i \cdot (1 - e^{-\Delta t / \tau _\mathit{rec} }) \cdot \frac{ 1 - e^{-\big( \mathit{OV(t_\mathit{dXT})}- V_L(t_\mathit{dXT}) \big) / V_0}} {1 - e^{-\mathit{OV}_0 / V_0}}
 \end{equation}
 where $\Delta t$ is the difference of $t_\mathit{dXT}$ and the preceding entry in the Geiger array with a finite amplitude.
 The term $1 - e^{-\Delta t / \tau _\mathit{rec}}$ takes the reduction in discharge probability of the delayed cross-talk discharge into account.
 The amplitude of the cross-talk is obtained by replacing $t_\mathit{pXT}$ by $t_\mathit{dXT}$ in Eq.\,\ref{eq:AXT}.

 The prompt and delayed cross-talk contributions from the pixel under consideration to the remaining pixels of the SiPM proceeds in an identical way, only the simulated Geiger Array of this pixel is used.
 All four cross-talk contributions are appended to the Geiger Array of the pixel under consideration.

 In order to simulate the data from Ref.\,\cite{Kotera:2016}, where the SiPM current is integrated in a gate of $t_\mathit{gate} = 150$\,ns, the integrated charge of a single pixel for a single event normalised to the nominal gain is:
 \begin{equation}\label{eq:Q1}
   Q_1 = \left( \sum _i A_i \cdot \int _{t_\mathit{tgate}} I(t - t_i)\,\mathrm{d}t \right) \cdot  Gauss(1,\sigma _0),
 \end{equation}
 where $I(t)$ is the SiPM current pulse from Eq.\,\ref{eq:SiPMpuls}, and $\mathit{Gauss} (1,\sigma_1)$ is a normally distributed random number with mean 1 and \emph{rms} $\sigma_0$, which accounts for the electronics noise.
 From many simulated single-pixel events the first moment, and the second and third central moments are calculated, and by multiplying with $N_\mathit{pix}$, the moments for the entire SiPM are obtained.

 \begin{figure}[!ht]
   \centering
   \begin{subfigure}[a]{0.5\textwidth}
    \includegraphics[width=\textwidth]{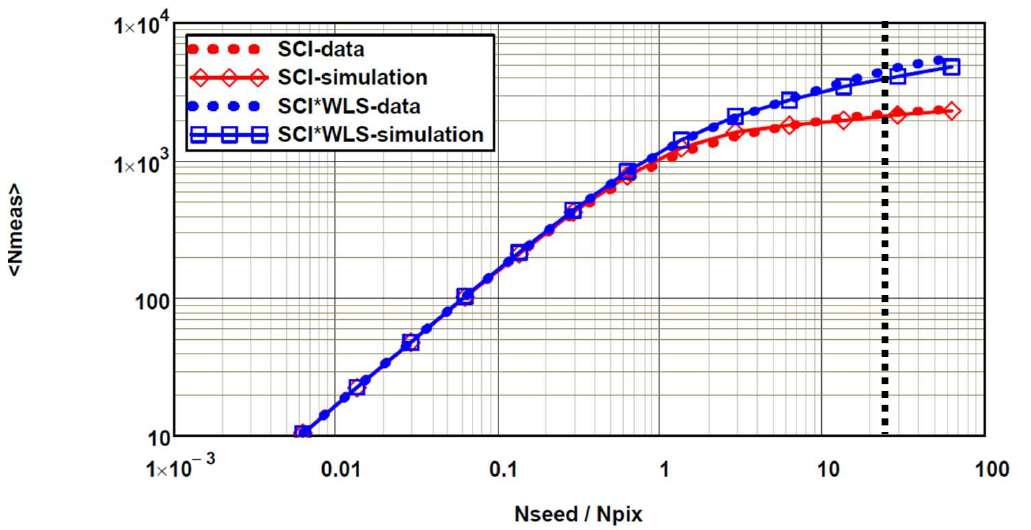}
    \caption{ }
    \label{fig:NsimKotera}
   \end{subfigure}%
    ~
   \begin{subfigure}[a]{0.5\textwidth}
    \includegraphics[width=\textwidth]{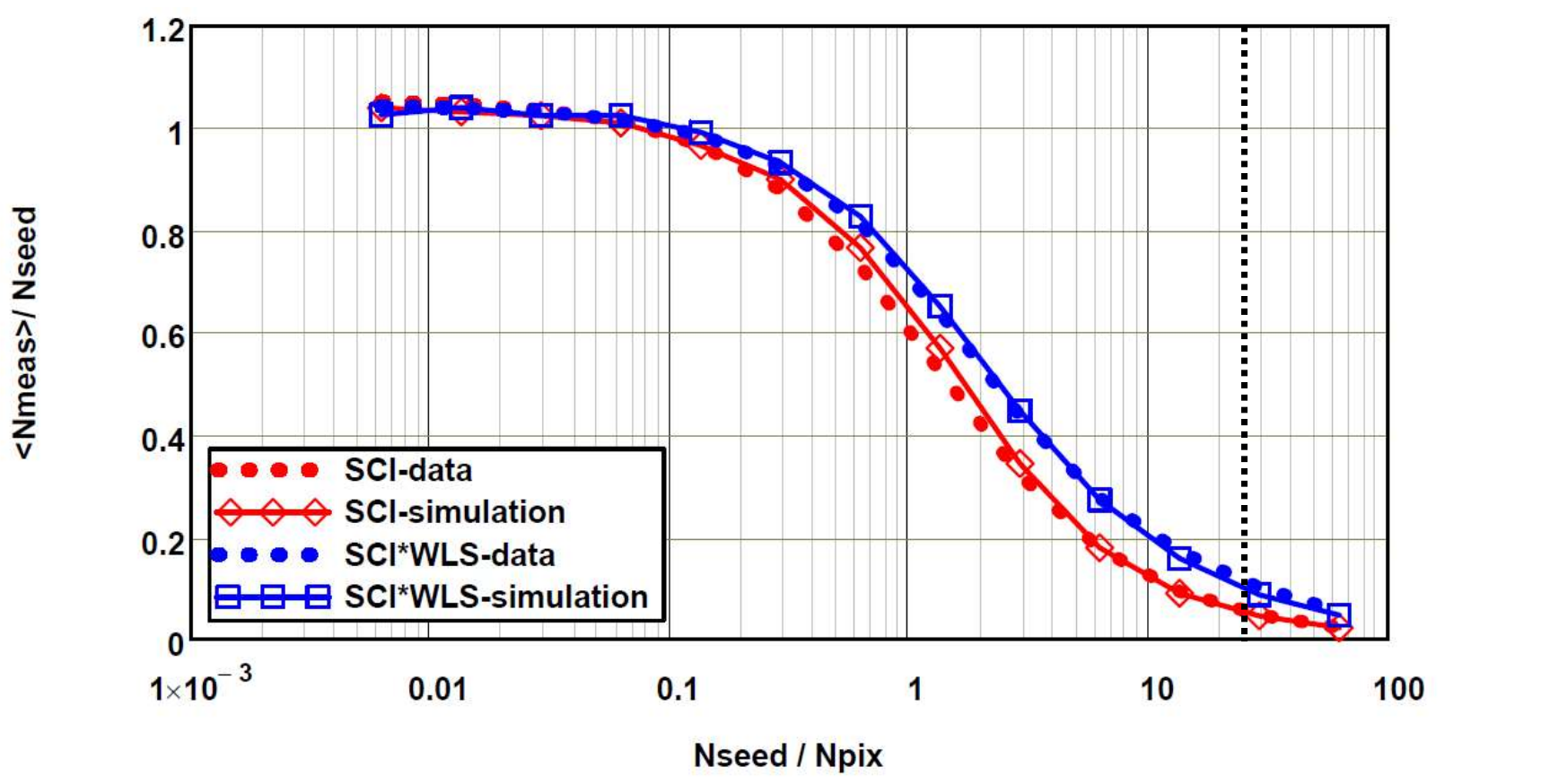}
    \caption{ }
    \label{fig:RsimKotera}
   \end{subfigure}%
   \caption{Comparison of the simulated response to the experimental results of Ref.\,\cite{Kotera:2016} for the setups \emph{SCI} and \emph{SCI*WLS} as a function of $N_\mathit{seed}/N_\mathit{pix}$, where $N_\mathit{seed}$ is the number of Geiger discharges in the absence of saturation effects.
   The comparison is made to the fits shown as continuous lines in Fig.\,\ref{fig:Kotera}:
   (a) The mean response, $N_\mathit{meas}$, in units of the charge of a single Geiger discharge at the nominal over-voltag, and
   (b) the ratio $N_\mathit{meas}/N_\mathit{seed}$, the signal reduction relative to no saturation.
   The vertical dotted lines show the approximate maximum $N_\mathit{seed}$\,values of the measurements.
   The simulations extend up to $N_\mathit{seed} = 10^5$, which corresponds to 62.5 potential Geiger discharges per pixel and light pulse for a SiPM with $N_\mathit{pix} = 1600$.
   }
  \label{fig:simKotera}
 \end{figure}

 Figure\,\ref{fig:simKotera} compares the mean response from the simulation to the fit to the data of Ref.\,\cite{Kotera:2016}, shown by the continuous lines of Fig.\,\ref{fig:Kotera}.
  To account for the contributions of after-pulses and cross-talk, \emph{ADC} has been multiplied by 1.045 to obtain $N_\mathit{seed}$.
 It can be seen that the simulation describes the experimental data within $\pm \,5$\,\%.
 The biggest deviation of $+5$\,\% between the simulation and the data for \emph{SCI} is in the region  $N_\mathit{seed} = ( 1 - 5 ) \cdot N_\mathit{pix}$, whereas for \emph{SCI*WLS} it is $-5$\,\% at the maximum $N_\mathit{seed}$\,values of the measurements.

 These discrepancies could not be removed by changing the parameters characterising the light pulses and the SiPM.
 This shows that the mean response is quite insensitive to the correlated change of various parameters, from which one can conclude that it is not possible to obtain precise information on these parameters from saturation measurements.

 \section{Comparison of the simulations to further experimental data}
  \label{sect:otherdata }

 In this section the results from the simulation are compared to the following published results:
  \begin{enumerate}
    \item The data from Ref.\,\cite{Weitzel:2019}, where the saturation properties for SiPMs with 100, 400, 1600 and 2668 pixels were investigated with sub-nanosecond light pulses of different intensities.
    \item The data from Ref.\,\cite{Bretz:2016}, where the saturation properties for SiPMs exposed to light pulses of different intensities and pulse lengths of 12, 35 and 73\,ns were investigated.
  \end{enumerate}

  \subsection{
  SiPMs with different number of pixels illuminated with sub-nanosecond light pulses }
   \label{subsect:Weitzel}

\begin{figure}[!ht]
   \centering
   \begin{subfigure}[a]{0.5\textwidth}
    \includegraphics[width=\textwidth]{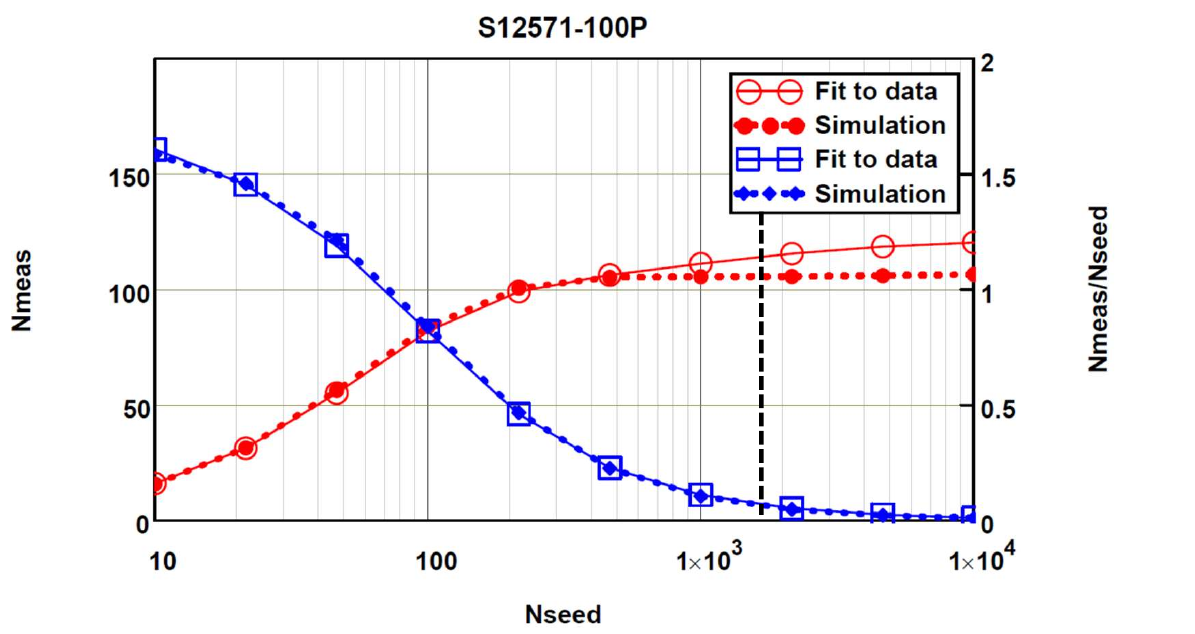}
    \caption{ }
    \label{fig:W-100P}
   \end{subfigure}%
    ~
   \begin{subfigure}[a]{0.5\textwidth}
    \includegraphics[width=\textwidth]{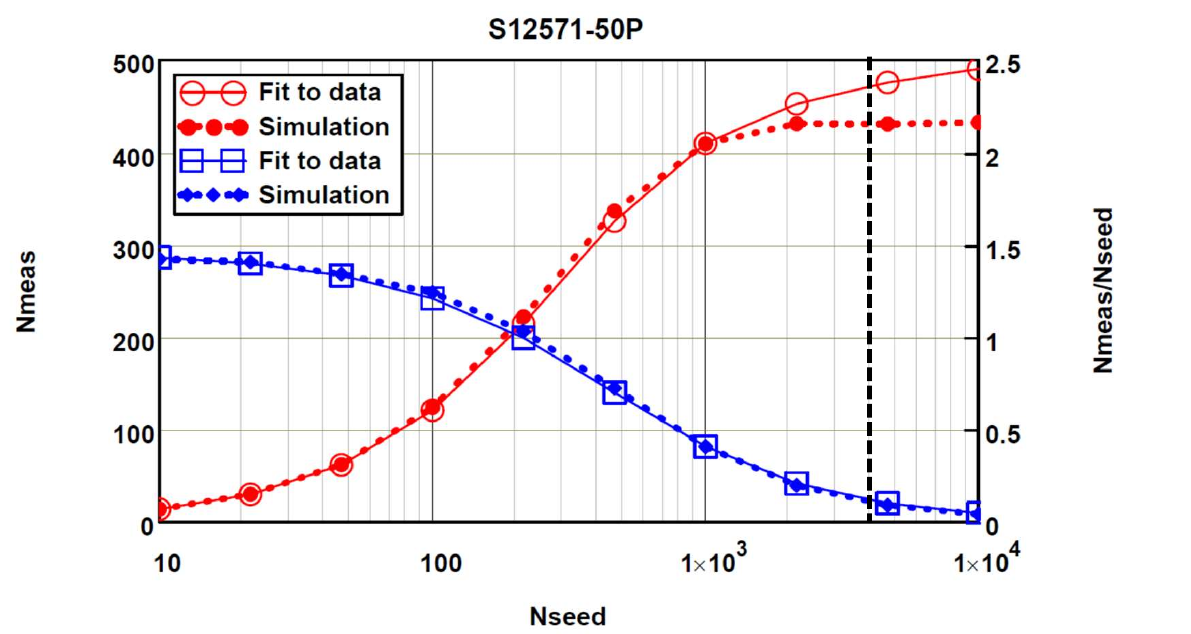}
    \caption{ }
    \label{fig:W-50P}
   \end{subfigure}%
   \newline
\begin{subfigure}[a]{0.5\textwidth}
    \includegraphics[width=\textwidth]{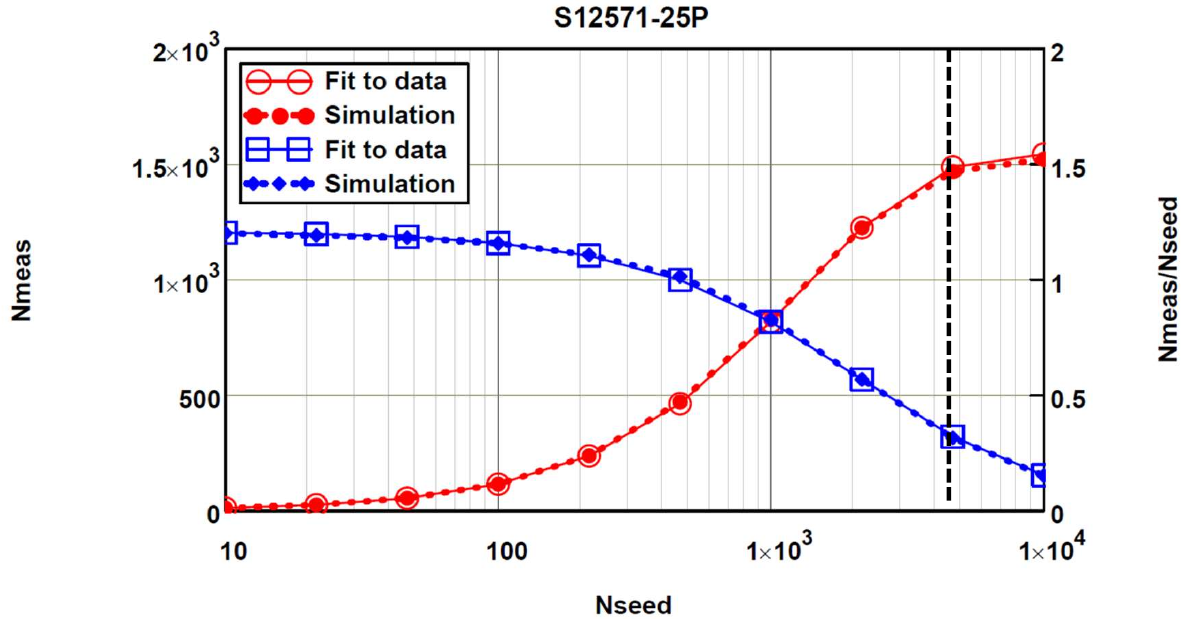}
    \caption{ }
    \label{fig:W-25P}
   \end{subfigure}%
    ~
   \begin{subfigure}[a]{0.5\textwidth}
    \includegraphics[width=\textwidth]{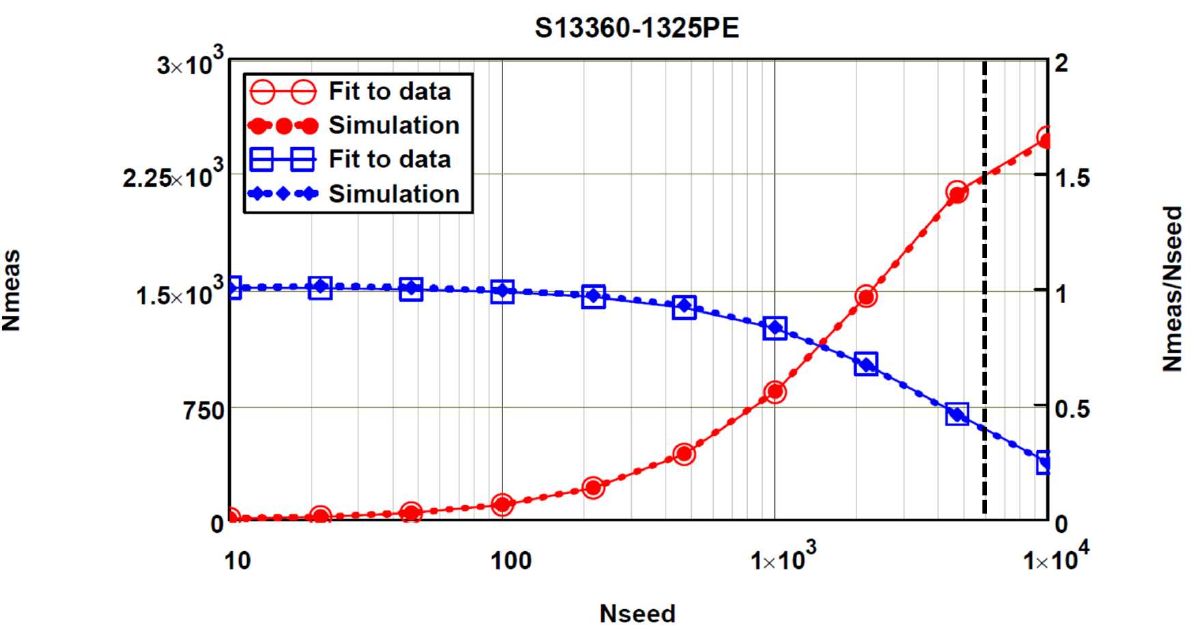}
    \caption{ }
    \label{fig:W-25PE}
   \end{subfigure}%
   \caption{Comparison of the simulation results to the data parametrisation from Ref.\,\cite{Weitzel:2019}.
   Shown are as a function of $N_\mathit{seed}$, the SiPM signal, $N_\mathit{meas}$, in units of the gain at the nominal over-voltage (boxes in blue, left scales), and $N_\mathit{meas}/N_\mathit{seed}$ (circles in red, right scales) for the SiPMs
   (a) S12751-100P with $N_\mathit{pix} = 100$,
   (b) S12751-50P with $N_\mathit{pix} = 400$,
   (c) S12751-25P with $N_\mathit{pix} = 1600$, and
   (d) S13360-1325PE with $N_\mathit{pix} = 2668$.
   The vertical dashed lines show the upper limit of the measurements.
   }
  \label{fig:W}
 \end{figure}

  Reference\,\cite{Weitzel:2019} reports saturation measurements by illuminating four SiPMs, with different number of pixels and pixel pitch, exposed to sub-nanosecond laser-light pulses of 467\,nm wavelength.
  SiPM parameters and the over-voltages at which the measurements were performed are shown in Table\,\ref{tab:W-SiPM}.
  The light intensity was monitored by a reference diode, and the SiPM signal was read out by the charge to digital converter QDC-V965A from CAEN.
  The gate width of $ \leq 100$\,ns was adapted to the width of the SiPM pulse, however the values are not given.
  The cross-talk parameter, $\mu _c$, which is the ratio of the measured signal to the signal from the photon-induced primary Geiger discharges, $N_\mathit{seed}$, in the linear range, was determined in a separate measurement discussed in Ref.\,\cite{Bauss:2016}.
  The data are presented as $N_\mathit{fired}$, which is called $N_\mathit{meas}$ in this paper, as a function of $N_\mathit{seed}$.
  A function with three free parameters is fitted to the experimental data, which is used in the present paper for the comparison to the simulations.
  Figure\,\ref{fig:W} shows the fit results as continuous lines connecting the open circles.
  As shown in Ref.\,\cite{Krause:2019}, the fitted function describes the data typically within $\pm 2$\,\%, however, for the SiPM with 100\,$\upmu$m pitch the fit deviates from the data by $-10$\,\% around $N_\mathit{seed} \approx 20$, and for the SiPM with 50\,$\upmu$m pitch by $+3$\,\% at $N_\mathit{seed} \approx 300$.

 \begin{table}[!ht]
  \caption{SiPMs used for the saturation measurements of Ref.\,\cite{Weitzel:2019}.
  The parameter $\mu _c$ accounts for after-pulses and cross-talk.
  For the linear range of the SiPMs $\mu _c = N_\mathit{meas}/N_\mathit{seed}$, the ratio of the signal with to the signal without after-pulses and cross-talk.
  }
  \label{tab:W-SiPM}
   \centering
    \begin{tabular}{c|r|c|c|c}
     Model & $N_\mathit{pix}$ & pitch [$\upmu$m] & $\mathit{OV}_0$ [V] & $\mu_c$ \\
   \hline
     S12571-100P & 100 & 100 & 1.66 & 1.89\\
     S12571-50P & 400 & 50 & 2.46 & 1.47 \\
     S12571-25P & 1600 & 25 & 3.31 & 1.36 \\
     S13360-1325PE & 2668 & 25 & 4.34 & 1.01 \\
\end{tabular}
\end{table}

 In the simulation for the time distribution of the primary Geiger discharges from photons a Gauss distribution with $\sigma = 150$\,ps has been used.
 This is significantly larger than the 50 to 100\,ps width of the light pulse from the laser.
 The additional width takes into account the time fluctuations in the build-up time of Geiger discharges.
 For the delay between the prompt cross-talk and the primary discharge a delay of 100\,ps has been introduced.
 The propagation time of a photon from the primary discharge to a neighbouring pixel is well below 1\,ps, however, according to Ref.\,\cite{Windischhofer:2023}, the build-up time for a Geiger discharge is of order 100\,ps.
 As a result, the step of the over-voltage and of the voltage drop over the load resistor shown in Fig.\,\ref{fig:OV} and Fig.\,\ref{fig:VL} are less abrupt, and the description of the data is improved.

 The SiPM parameters required for the simulations are very much correlated: Different sets of parameters provide equally good descriptions of the data.
 Therefore, the values of the individual parameters of Table\,\ref{tab:Parameters} have little meaning, and are not given.
 However, it is noted that to describe the data for the $25\,\upmu$m SiPMs require a value of the quenching resistance $R_q \gtrsim 1$\,M$\Omega$.

 Fig.\,\ref{fig:W} compares the simulation results to the data as a function of $N_\mathit{seed}$:
 $N_\mathit{meas}$ boxes in blue, with the scale on the left, and the signal reduction due to saturation effects, $N_\mathit{meas}/N_\mathit{seed}$, circles in red, with the scale on the right.
 Overall, the simulation describes the data within their uncertainties with the exception of the  $100\,\upmu$m SiPM and the $50\,\upmu$m SiPM for $n_\mathit{seed} = N_\mathit{seed}/N_\mathit{pix} \gtrsim 4$.
 Whereas the data show a further increase of $N_\mathit{meas}$, the simulations are essentially constant.
 According to Poisson statistics, the fraction of pixels without Geiger discharges is $e^{-n_\mathit{seed}}$.
 Assuming that the signal increase is given by the reduction of pixels without discharges, an
 increase between $n_\mathit{seed} = 4$ and $n_\mathit{seed} \rightarrow \infty$ of $e^{-4} \approx 2$\,\% is expected, whereas an increase by about 13\,\% is  observed.
 We do not understand this difference.
 A possible explanation could be that the signal from two or more Geiger discharges occurring simultaneously in a pixel is bigger than the signal from a single discharge.


 To summarise this subsection:
 For four SiPMs with $N_\mathit{pix} = $\,100, 400, 1600 and 2668 and pixel pitches of 100, 50, 25, and 25\,$\upmu$m, the simulation program is able to accurately describe the experimental results of Ref.\,\cite{Weitzel:2019} up to photon intensities corresponding to an average number of about four potential Geiger discharges per pixel.
 This corresponds to a signal reduction due to saturation effects by about a factor four.
 For more than four potential Geiger discharges per pixel, where data are available only for the SiPMs with $N_\mathit{pix} = $\,100 and 400, the measured signal still increases, whereas the simulation predicts a constant value.
 This observation is not understood and deserves further studies.

   \subsection{Single SiPM illuminated with photon pulses of different durations}
    \label{subsect:Bretz}

 Reference \cite{Bretz:2016} reports saturation measurements by illuminating three SiPM types at different over-voltages for light pulses with photon numbers over the sensitive area, $N_\gamma $ of up to  $6 \times 10^6$ and pulse widths of 12, 35 and 73\,ns.
 In the present paper we only compare the simulations to the data taken with the KETEK SiPM PM3350 which has $N_\mathit{pix} = 3600$ pixels of $50\,\upmu$m\,pitch, and was operated at nominal over-voltages $\mathit{OV}_0 = 2.1,$ 3.1 and 4.1\,V.
 As light source an LED emitting light at 480\,nm was used, which was driven by a custom-made LED pulser with computer controlled amplitude and pulse width.
 The light was send to an integrating sphere with 2 exits connected to optical fibers;
 one for illuminating the SiPM and the other for illumination a calibrated photodiode.
 The SiPM was operated at $\,0^\circ$C.
 For the read out a QDC V965 from CAEN with 50\,$\Omega$ input resistance has been used.
 The gate length was 400\,ns.
 The SiPM has been characterised and the parameters are shown in Table\,\ref{tab:ParBretz}.

 \begin{table}[!ht]
  \caption{Parameters of the KETEK PM3350 SiPM taken from Table 2 of Ref.\,\cite{Bretz:2016}to the left of the double line, and parameters used for the simulation to the right.
  \emph{XT} and \emph{Ap} are the increase in measured charge from cross-talk and after-pulses.
  For the explanation of the remaining symbols see Table\,\ref{tab:Parameters}.
  }
  \label{tab:ParBretz}
  \centering
  \begin{tabular}{c|c|c|c|c||c|c|c|c|c}
   $ \mathit{OV}_0$ & \emph{DCR} & \emph{XT} & $p_\mathit{Ap}$  & $\tau_s$ & $p_\mathit{Ap}$ & $\tau_\mathit{Ap}$ & $p_\mathit{pXT}$ & $p_\mathit{dXT}$ & $\tau_\mathit{dXT}$ \\
   $[\mathrm{V}]$ & [kHz] & [\%] & [\%] & [ns] & [\%] & [ns] & [\%] & [\%]& [ns] \\
 \hline
   2.1 & 110 & 3  & < 9 & 83 & 2   & 10 & 2 & 2 & 20 \\
   3.1 & 200 & 6  & < 9 & 83 & 2.5 & 10 & 3 & 3 & 20 \\
   4.1 & 330 & 10 & < 9 & 83 & 3   & 10 & 5 & 5 & 20 \\
   \hline
  \end{tabular}
 \end{table}

 The measurements for the 12\,ns pulse-width data extend up to $N_\gamma = 1.2 \times 10^5$, the ones for 35\,ns to $5 \times 10^5$ and the ones for 73\,ns to $5 \times 10^5$ with the exception of the 4.1\,V data, which stop at $1.2 \times 10^5$.
 The data of Fig.\,11 of Ref.\,\cite{Bretz:2016}, where $N_\mathit{fired}$, called $N_\mathit{meas}$ in this paper, as a function of $N_\gamma$ is shown, have been provided by M.\,Lauscher.
 As before, the measured charge in units of the nominal charge of a single Geiger discharge is called $N_\mathit{meas}$.
 It turns out that the data for low $N_\gamma $ values have significant uncertainties.
 Therefore, for  $N_\gamma < 5 \times 10^4$, where the non-linearity is small and independent of over-voltage and pulse length, the results of the fit of Eq.\,5.1
 of Ref.\,\cite{Bretz:2016} to the data has been used.

 For the simulation $N_\mathit{seed} = \mathit{pde} \cdot N_\gamma$ is required.
 The photon-detection-efficiency can be estimated by $ \mathit{pde} = N_\mathit{meas}/\left(N_\gamma \cdot (1 + \mathit{XT}) \cdot ( 1 + \mathit{Ap} )\right) $, where here $N_\mathit{meas}$ is the measured signal with the effects of \emph{DCR} subtracted; \emph{XT} is the increase of the signal due to cross-talk and \emph{Ap} due to after-pulses.
 This relation can be derived in the following way:
 In the linear region the measurements determine the slope $\mathit{d}N_\mathit{meas} / \mathit{d}N_\gamma =  N_\mathit{meas}/N_\gamma$.
  With $N_\mathit{meas} = N_\mathit{seed} \cdot (1 + \mathit{XT}) \cdot (1 + \mathit{Ap})$ and using $N_\gamma = N_\mathit{seed} /\mathit{pde}$, above formula is obtained.

  \begin{figure}[!ht]
   \centering
   \begin{subfigure}[a]{0.5\textwidth}
    \includegraphics[width=\textwidth]{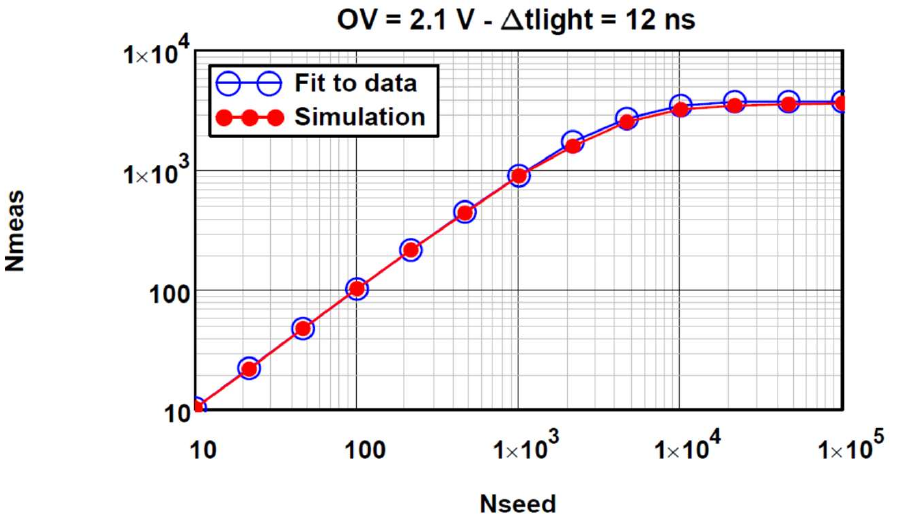}
    \caption{ }
    \label{fig:N2p1V12ns}
   \end{subfigure}%
    ~
   \begin{subfigure}[a]{0.5\textwidth}
    \includegraphics[width=\textwidth]{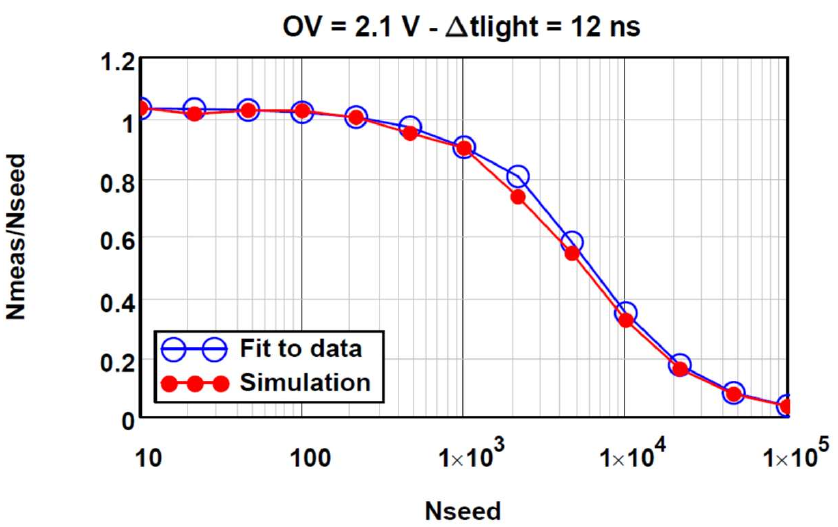}
    \caption{ }
    \label{fig:R2p1V12ns}
   \end{subfigure}%
   \newline
\begin{subfigure}[a]{0.5\textwidth}
    \includegraphics[width=\textwidth]{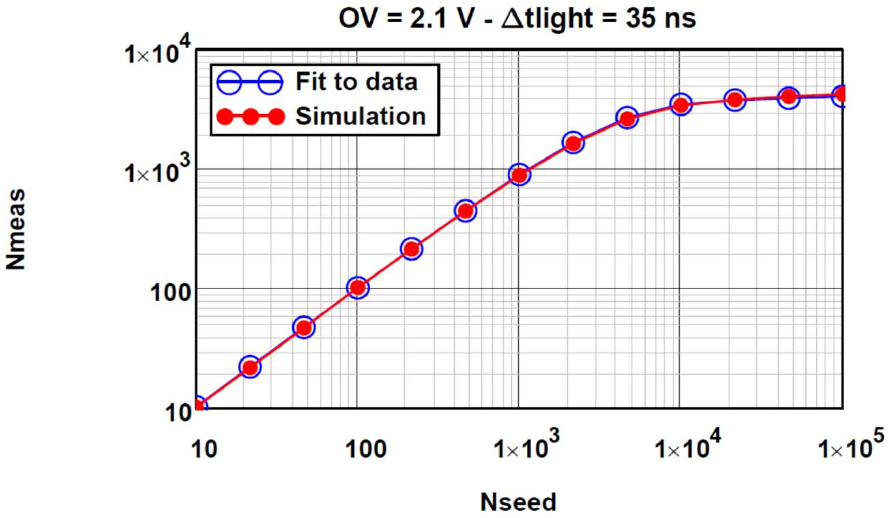}
    \caption{ }
    \label{fig:W-N2p1V35ns}
   \end{subfigure}%
    ~
   \begin{subfigure}[a]{0.5\textwidth}
    \includegraphics[width=\textwidth]{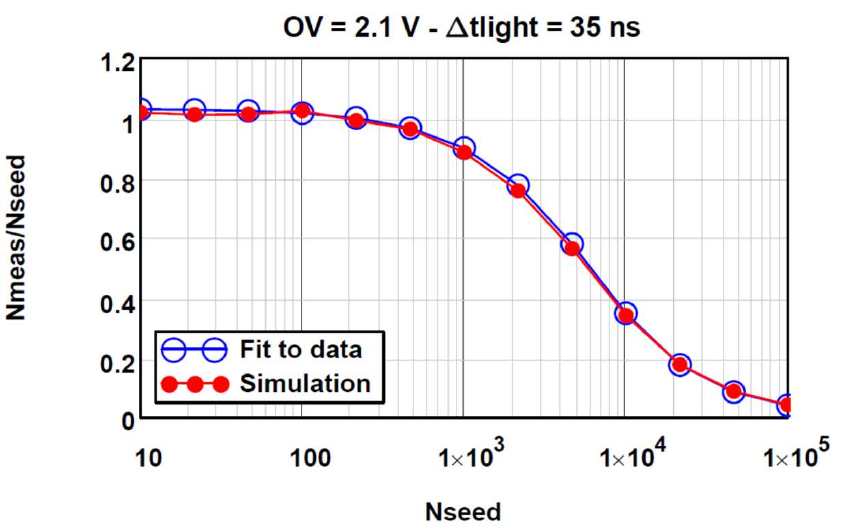}
    \caption{ }
    \label{fig:R2p1V73ns}
   \end{subfigure}%
   \newline
\begin{subfigure}[a]{0.5\textwidth}
    \includegraphics[width=\textwidth]{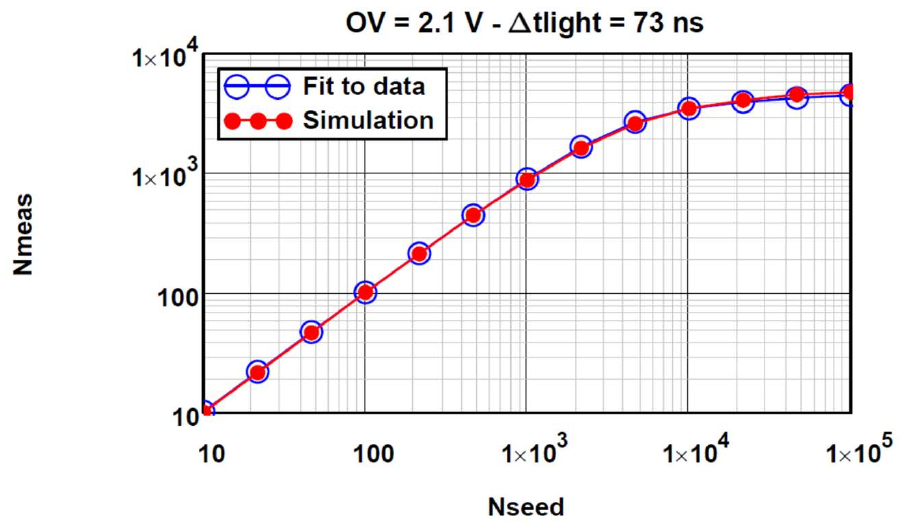}
    \caption{ }
    \label{fig:N2p1V73ns}
   \end{subfigure}%
    ~
   \begin{subfigure}[a]{0.5\textwidth}
    \includegraphics[width=\textwidth]{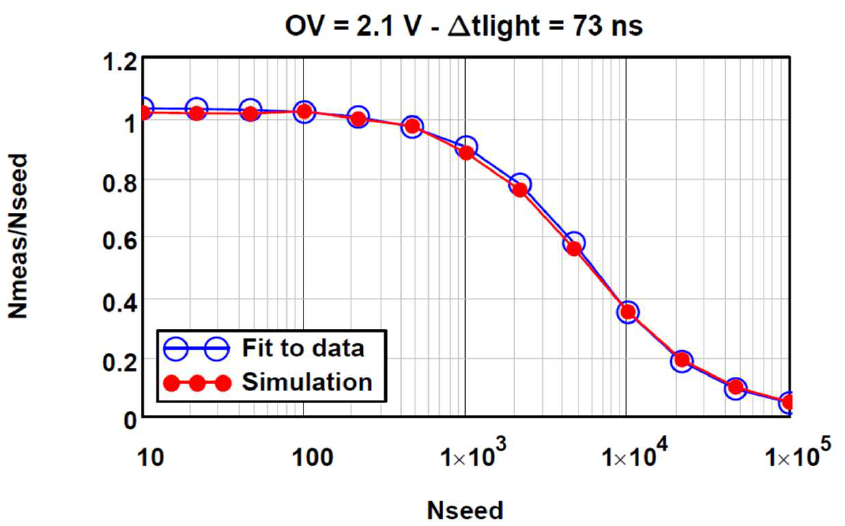}
    \caption{ }
    \label{fig:R2p1V73ns}
   \end{subfigure}%
   \caption{Comparison of the simulation results to the data of Ref.\,\cite{Bretz:2016} at an over-voltage $\mathit{OV}_0 = 2.1$\,V.
   Shown are as a function of $N_\mathit{seed}$ the SiPM signal, $N_\mathit{meas}$, in units of the gain at $\mathit{OV}_0$  (a, c, e), and the ratio $N_\mathit{meas}/N_\mathit{seed}$ (b, d, f).
   The values of the over-voltage, $\mathit{OV}_0$, and of the widths of the light pulses, $\Delta t_\mathit{light}$, are given on the top of the sub-figures.
   }
  \label{fig:N2p1V}
 \end{figure}

  \begin{figure}[!ht]
   \centering
   \begin{subfigure}[a]{0.5\textwidth}
    \includegraphics[width=\textwidth]{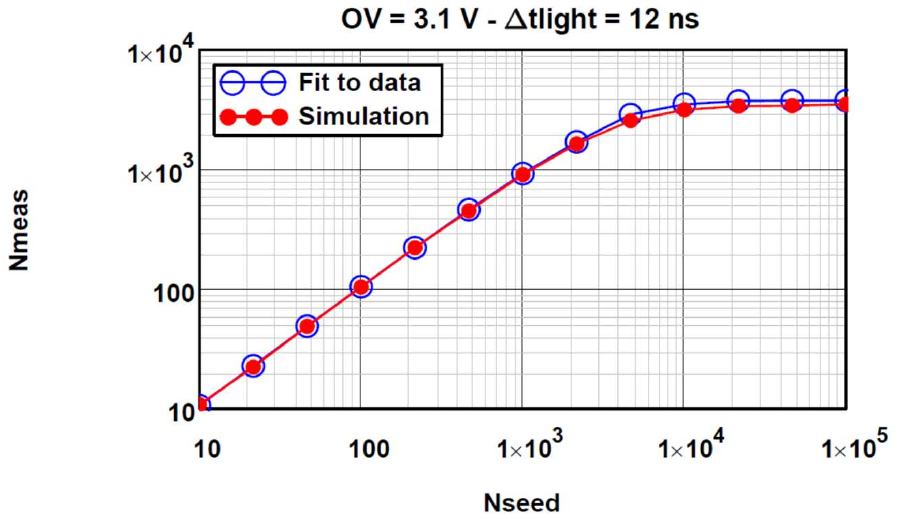}
    \caption{ }
    \label{fig:N3p1V12ns}
   \end{subfigure}%
    ~
   \begin{subfigure}[a]{0.5\textwidth}
    \includegraphics[width=\textwidth]{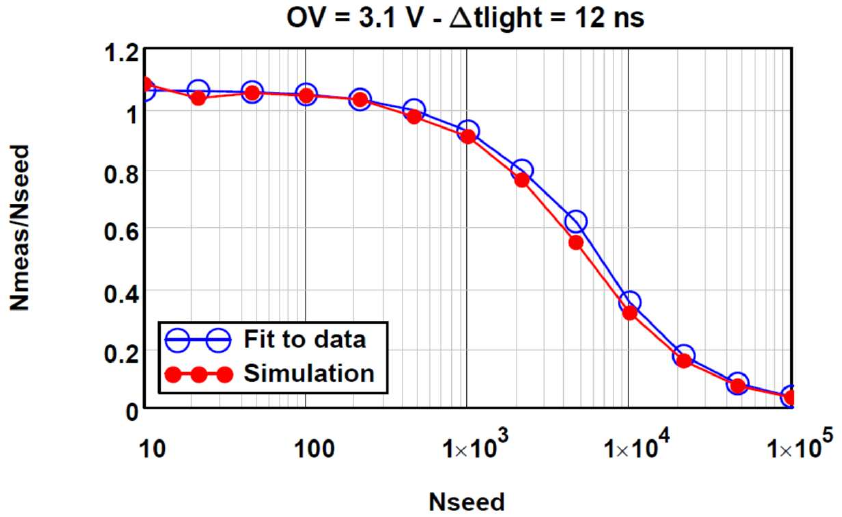}
    \caption{ }
    \label{fig:R3p1V12ns}
   \end{subfigure}%
   \newline
\begin{subfigure}[a]{0.5\textwidth}
    \includegraphics[width=\textwidth]{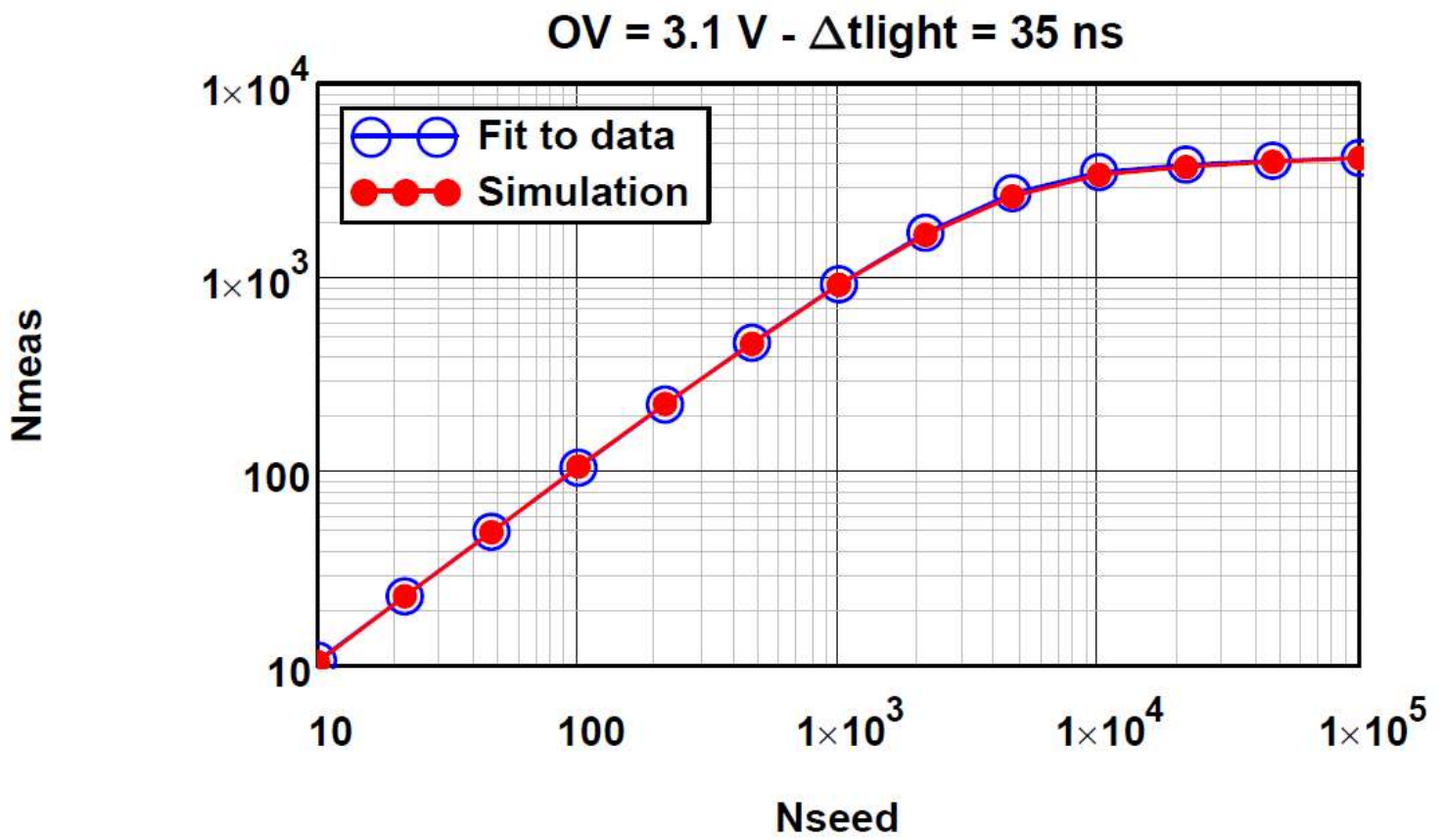}
    \caption{ }
    \label{fig:W-N3p1V35ns}
   \end{subfigure}%
    ~
   \begin{subfigure}[a]{0.5\textwidth}
    \includegraphics[width=\textwidth]{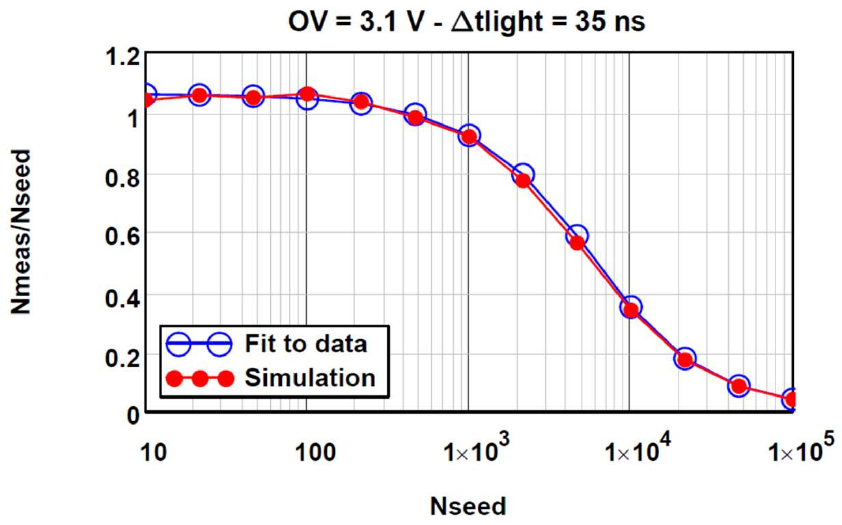}
    \caption{ }
    \label{fig:R3p1V73ns}
   \end{subfigure}%
   \newline
\begin{subfigure}[a]{0.5\textwidth}
    \includegraphics[width=\textwidth]{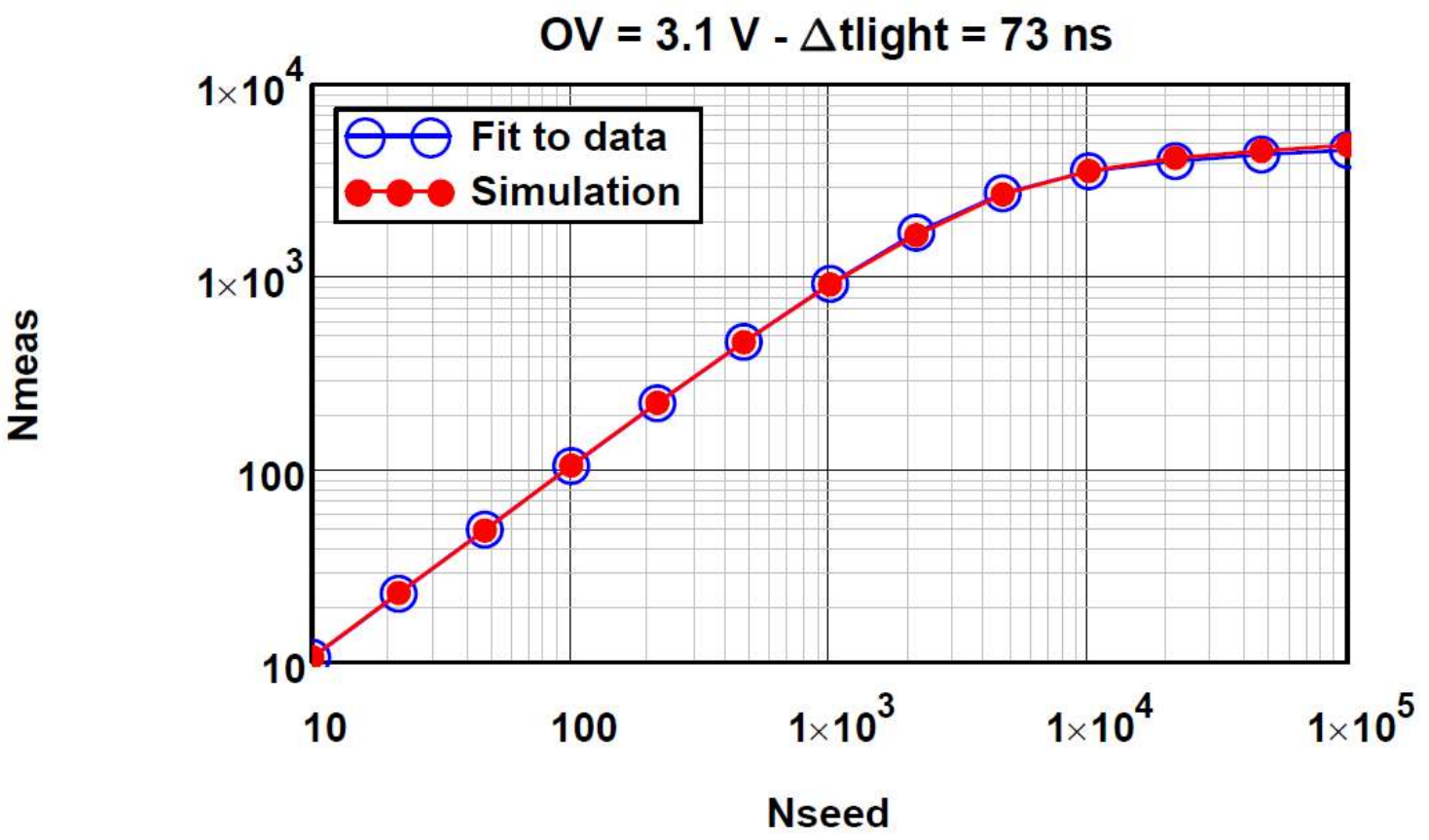}
    \caption{ }
    \label{fig:N3p1V73ns}
   \end{subfigure}%
    ~
   \begin{subfigure}[a]{0.5\textwidth}
    \includegraphics[width=\textwidth]{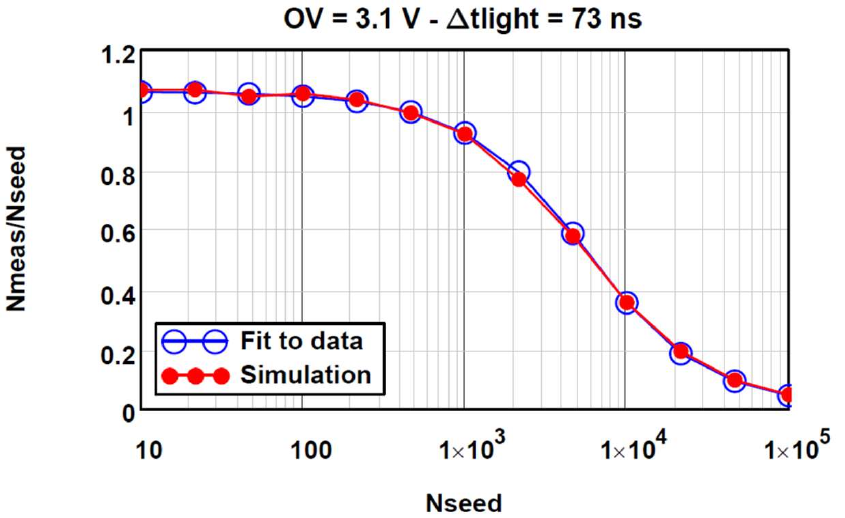}
    \caption{ }
    \label{fig:R3p1V73ns}
   \end{subfigure}%
   \caption{Same as Fig.\,\ref{fig:N2p1V} for $\mathit{OV}_0 = 3.1$\,V.
   }
  \label{fig:N3p1V}
 \end{figure}

  \begin{figure}[!ht]
   \centering
   \begin{subfigure}[a]{0.5\textwidth}
    \includegraphics[width=\textwidth]{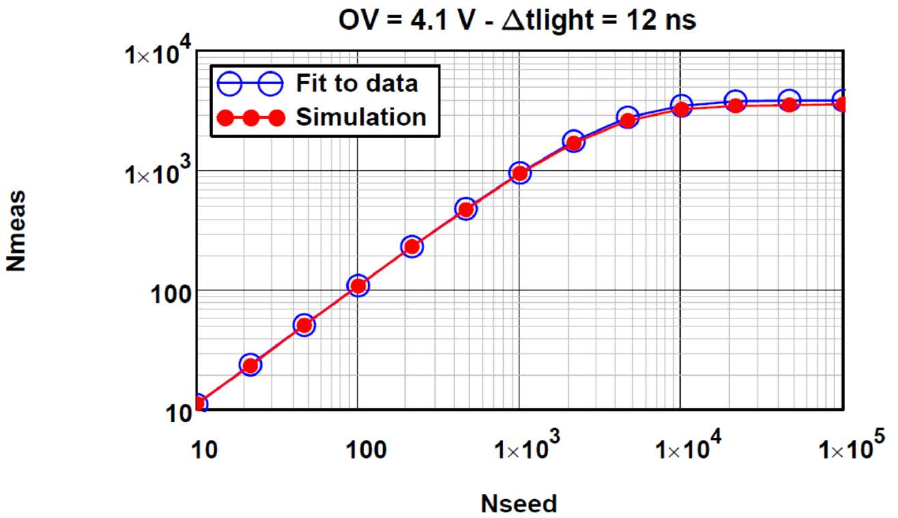}
    \caption{ }
    \label{fig:N4p1V12ns}
   \end{subfigure}%
    ~
   \begin{subfigure}[a]{0.5\textwidth}
    \includegraphics[width=\textwidth]{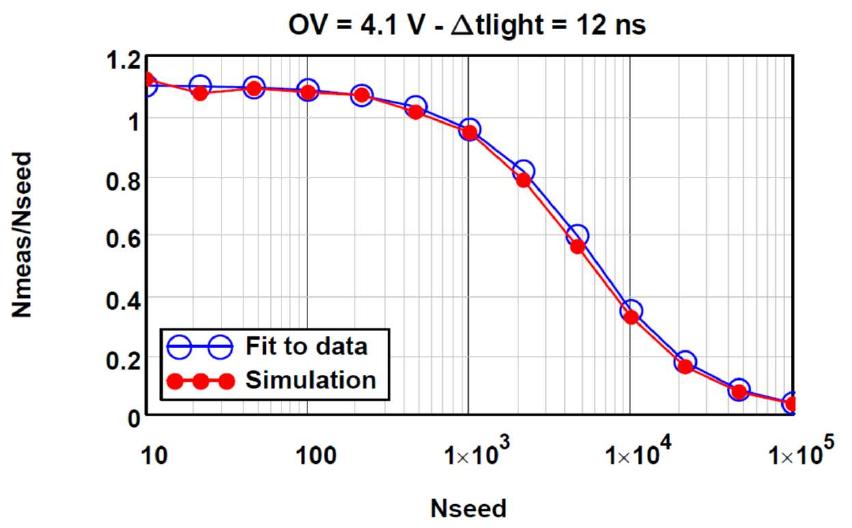}
    \caption{ }
    \label{fig:R4p1V12ns}
   \end{subfigure}%
   \newline
\begin{subfigure}[a]{0.5\textwidth}
    \includegraphics[width=\textwidth]{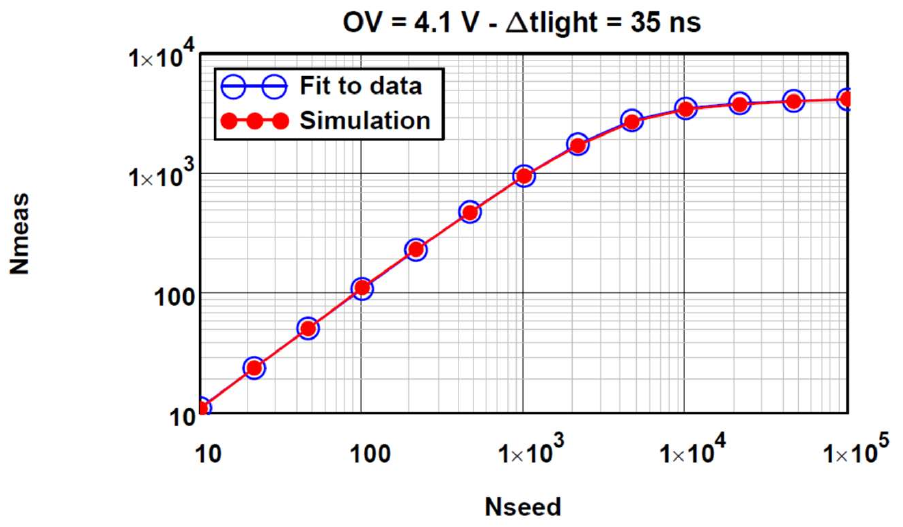}
    \caption{ }
    \label{fig:W-N4p1V35ns}
   \end{subfigure}%
    ~
   \begin{subfigure}[a]{0.5\textwidth}
    \includegraphics[width=\textwidth]{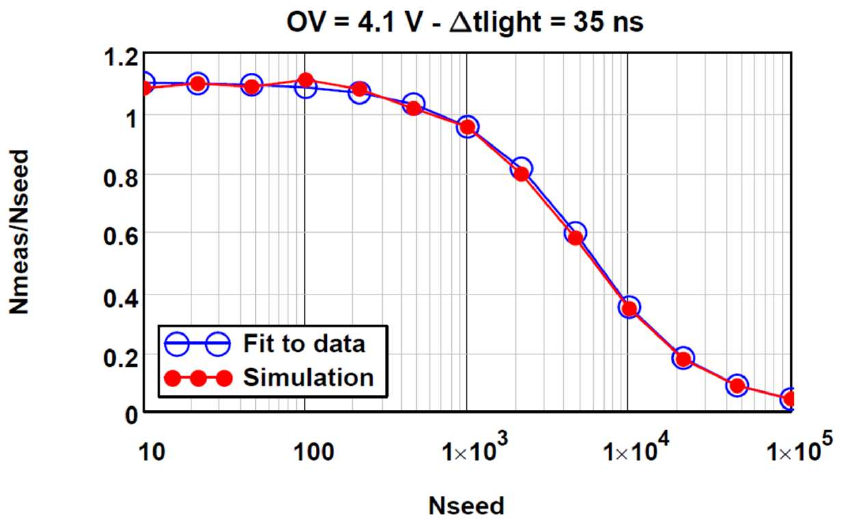}
    \caption{ }
    \label{fig:R4p1V73ns}
   \end{subfigure}%
   \newline
\begin{subfigure}[a]{0.5\textwidth}
    \includegraphics[width=\textwidth]{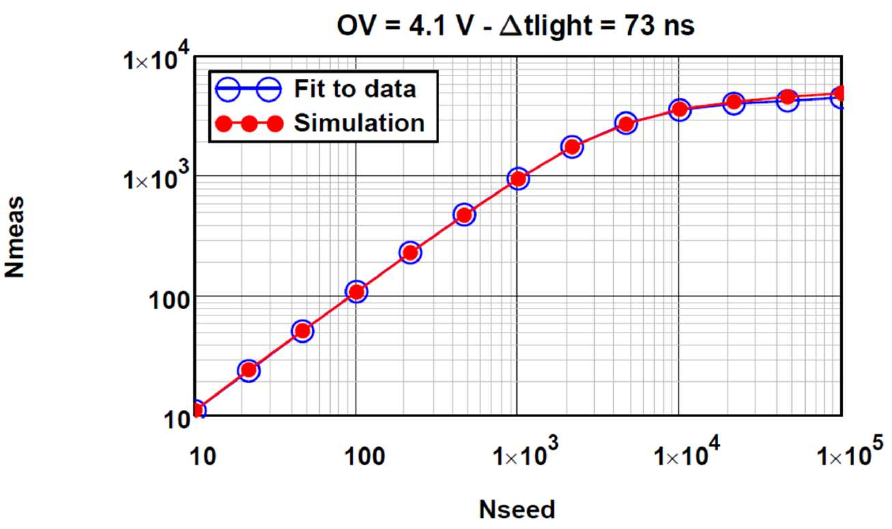}
    \caption{ }
    \label{fig:N4p1V73ns}
   \end{subfigure}%
    ~
   \begin{subfigure}[a]{0.5\textwidth}
    \includegraphics[width=\textwidth]{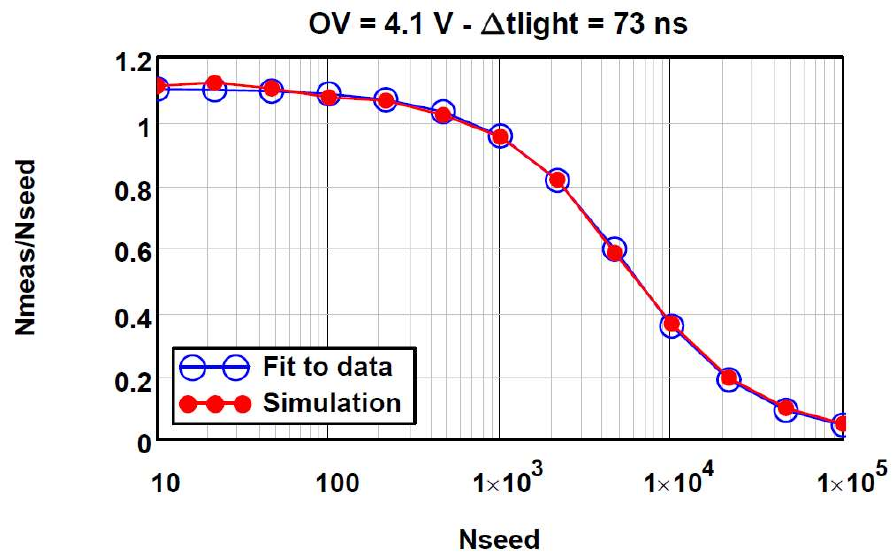}
    \caption{ }
    \label{fig:R4p1V73ns}
   \end{subfigure}%
   \caption{Same as Fig.\,\ref{fig:N2p1V} for $\mathit{OV}_0 = 4.1$\,V.
   }
  \label{fig:N4p1V}
 \end{figure}

 The open circles of Figs.\,\ref{fig:N2p1V} to \ref{fig:N4p1V} show selected values of $N_\mathit{meas}$ as a function of $N_\mathit{seed}$ for the measurements.
 They are shown in double-logarithmic scale, so that the transition from the linear to the non-linear region, which is of interest for most applications, can be seen clearly.
 For the simulations the parameters given in the columns to the right of the double line of Table\,\ref{tab:ParBretz} and the value of \emph{DCR} have been used.
 The simulation has been performed between $-t_0 = -100$\,ns and $t_\mathit{gate}= 400$\,ns.
 The results are shown as filled circles.
 The left sides of the figures show $N_\mathit{meas}$, the simulated charge in units of the nominal charge of a single Geiger discharge, and the right sides the non-linearity $N_\mathit{meas}/N_\mathit{seed}$.

 Overall the simulations describe the experimental data to better than $\pm 5$\,\%, which is approximately the uncertainty of the data.
 An exception are the data for 12\,ns long light pulses in the region $N_\mathit{seed} = N_\mathit{pix} =3600$, where the simulation is about 10\,\% lower than the data.
 This discrepancy could not be removed by changing the parameters of the simulation, and is not understood.

\begin{figure}[!ht]
   \centering
   \begin{subfigure}[a]{0.5\textwidth}
    \includegraphics[width=\textwidth]{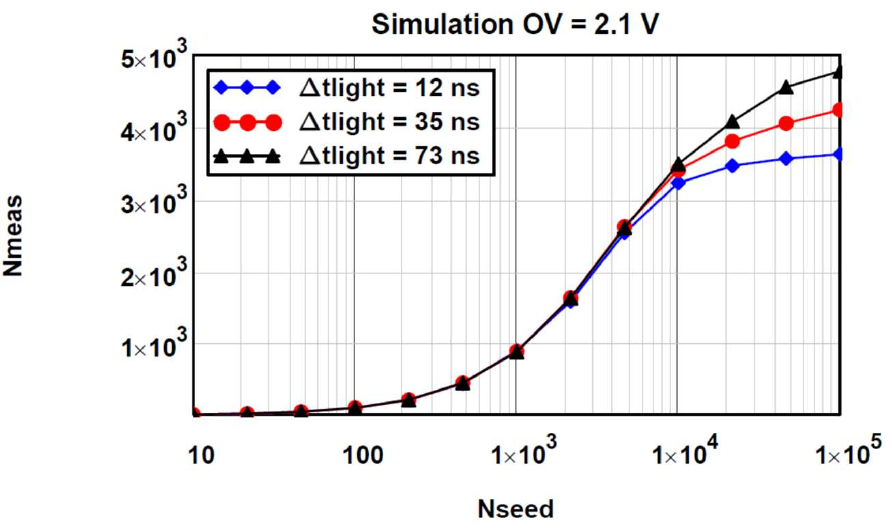}
    \caption{ }
    \label{fig:N2p1}
   \end{subfigure}%
    ~
   \begin{subfigure}[a]{0.5\textwidth}
    \includegraphics[width=\textwidth]{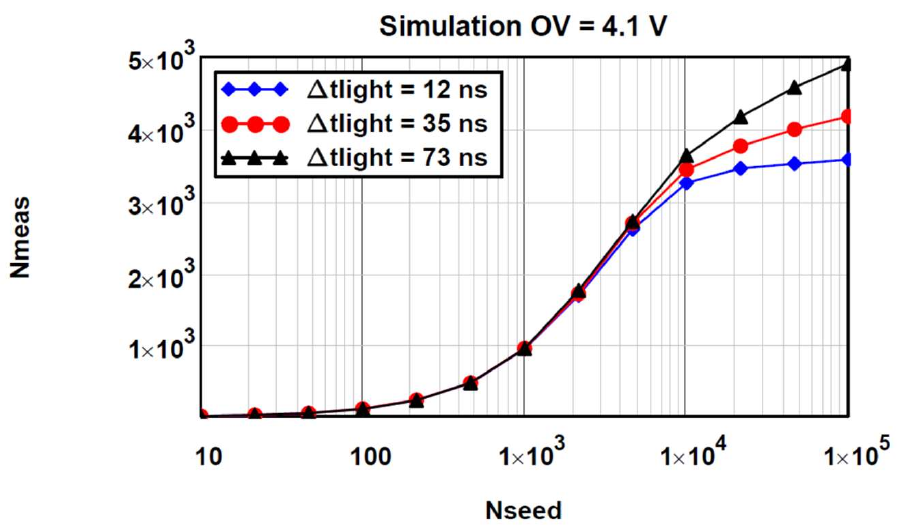}
    \caption{ }
    \label{fig:N4p1}
   \end{subfigure}%
   \caption{Simulated SiPM saturation for different widths of the light pulses, $\Delta t_\mathit{light}$,
   (a) for $\mathit{OV}_0 = 2.1$\,V, and
   (b) for $\mathit{OV}_0 = 4.1$\,V.
   }
  \label{fig:NV}
 \end{figure}

 To demonstrate the dependence of the SiPM saturation on the length of the light pulse, Fig.\,\ref{fig:NV} shows the simulated $N_\mathit{meas}$ at 2.1\,V and 4.1\,V for the light-pulse lengths of 12, 35 and 73\,ns, in linear scale.
 It can be seen that up to $N_\mathit{seed} = N_\mathit{pix} = 3600$, the SiPM response does not depend on the length of the light pulse, although the ratio $N_\mathit{meas}/N_\mathit{seed} \approx 0.7$ at $N_\mathit{seed} = 3600$ (right side of Figs.\,\ref{fig:N2p1V} to \ref{fig:N4p1V}) shows already a significant non-linearity.
 For higher $N_\mathit{seed}$\,values, the SiPM response strongly depends on the length of the light pulse:
 For $N_\mathit{seed} = 10^5$, i.e. an average of about 28 Geiger-discharge candidates per pixel,
 the ratios $N_\mathit{meas}/N_\mathit{pix}$ are approximately 1.0, 1.12, and 1.33, for light-pulse lengths of 12, 35 and 73\,ns, respectively.
 It is noted that these values are approximately independent of the over-voltage.

 To summarize this subsection:
 For a SiPM with $N_\mathit{pix} = 3600$ pixels with a pitch of $50\,\upmu$m, operated at over-voltages of 2.1, 3.1 and 4.1\,V, and illuminated with square light pulses of 12, 35 and 73\,ns duration and photon intensities resulting in  $N_\mathit{seed}$\,values up to $10^5$, the simulation program is able to describe the experimental data of Ref.\,\cite{Bretz:2016} within typically less than $\pm 5$\,\%.
 The experimental data and the simulations show
 that up to $N_\mathit{seed} \approx N_\mathit{pix}$, where about 30\,\% of the signal is lost due to saturation, the measured charge is independent of the duration of the light pulse,
 and that for higher $N_\mathit{seed}$\,values, the saturation decreases significantly with increasing pulse duration.


 \newpage

 \section{Summary}
  \label{sect:summary}

  A Monte Carlo program is presented which simulates the response of SiPMs in the nonlinear regime, where the number of Geiger discharges from photons and/or from dark counts in the time interval given by the pulse shape of a single Geiger discharge approaches or exceeds the number of SiPM pixels.
 The model includes the effects of after-pulses, of prompt and delayed cross-talk and of the voltage drop over the load resistance of the readout electronics.

  The results of the simulation program are compared to three publications which studied the response of SiPMs as a function of the light intensity for different shapes of the light pulse, different numbers and dimensions of the SiPM pixels and different applied over-voltages.
 The experimental data extend to very high saturation values: up to $\approx 25$ potential Geiger discharges per SiPM pixel.
  The simulation describes the experimental data typically within $\pm\,5$\,\%.
 An exception is the saturation for sub-nanosecond laser light, for which the program predicts a constant saturation value, whereas the data show a further increase with light intensity.
  This could be explained if the response of two or more simultaneous Geiger discharges in a pixel result in a larger signal than of a single Geiger discharge.

 So far, the comparison of the simulation results to experimental data for saturation effects due to high dark-count rates from ambient light or radiation damage has not been made.
  No quantitative results for the effects of ambient light has been found in the literature, and for radiation-damaged SiPMs, it is not clear how much SiPM performance parameters like gain, after-pulses and cross-talk change with irradiation and temperature.
 Such studies, which are relevant for e.\,g. LIDAR, $\gamma $-ray cameras in astro-particle physics and calorimetry at colliders, should be performed, and it is expected that the simulation program will be useful for these investigations.

 \section*{Acknowledgements}
  \label{sect:Acknowledgement}

   The author thanks Markus Lauscher for providing the ASCII file with the content of the histograms of Fig.\,11 of Ref.\,\cite{Bretz:2016} and in his interest in the paper.
    The author is also grateful to Erika Garutti and J\"orn Schwandt for discussions and comments which resulted in improvements of the manuscript.


\end{document}